\documentclass[3p, 11pt, amsmath, amssymb]{elsarticle}
\usepackage{bm,amsfonts, mathtools}
\usepackage{graphicx,color, wasysym}
\usepackage{textcomp}
\usepackage{amsmath,amssymb,latexsym,epsfig}
\usepackage[british]{babel}
\usepackage{hyphenat}
\usepackage{appendix}
\usepackage{mathrsfs}
\usepackage{caption}
\usepackage{subcaption}

\DeclareMathOperator{\D}{d\!}
\DeclareMathOperator{\E}{e} 
\DeclareMathOperator{\I}{i}
   \DeclareMathOperator{\RE}{\mathfrak{Re}}
   \DeclareMathOperator{\IM}{\mathfrak{Im}}

\newtheorem{theorem}{Theorem}

\newtheorem{proposition}{Proposition}
\newtheorem{remark}{Remark}

\newtheorem{example}{Example}

\begin{document} 

\allowdisplaybreaks

\makeatletter

\title{Volterra-Prabhakar derivative of distributed order and some applications}

\author[rvt]{K.~G\'{o}rska\corref{cor1}}
\address[rvt]{H. Niewodnicza\'{n}ski Institute of Nuclear Physics, Polish Academy of Sciences, \\ 
ul.Eljasza-Radzikowskiego 152, PL 31342 Krak\'{o}w, Poland}
\cortext[cor1]{Corresponding author.}
\ead{katarzyna.gorska@ifj.edu.pl}

%\author[rvt]{A.~Horzela}
%\ead{andrzej.horzela@ifj.edu.pl}

\author[rvt]{T.~Pietrzak}
\ead{tobiasz.pietrzak@ifj.edu.pl}

\author[r3]{T. Sandev}
\address[r3]{Research Center for Computer Science and Information Technologies, Macedonian Academy of Sciences and Arts, \\Bul. Krste Misirkov 2, 1000 Skopje, Macedonia\\ Institute of Physics, Faculty of Natural Sciences and Mathematics, Ss. Cyril and Methodius University, \\ Arhimedova 3, 1000 Skopje, Macedonia \\ Institute of Physics \& Astronomy, University of Potsdam, D-14776 Potsdam-Golm, Germany}
\ead{trifce.sandev@manu.edu.mk}

\author[r2]{\v Z.~Tomovski}
\address[r2]{University of Ostrava, Faculty of Sciences, Department of Mathematics, \\ 30. Dubna 22701 03 Ostrava, Czech Republic}
\ead{zhivorad.tomovski@osu.cz}

\begin{abstract}
The paper studies the exact solution of two kinds of generalized Fokker-Planck equations in which the integral kernels are given either by the distributed order function $k_{1}(t) = \int_{0}^{1} t^{-\mu}/\Gamma(1- \mu) \D\mu$ or the distributed order Prabhakar function  $k_{2}(\alpha, \gamma; \lambda; t) = \int_{0}^{1} e^{-\gamma}_{\alpha, 1 - \mu}(\lambda; t) \D\mu$, where the Prabhakar function is denoted as $e^{-\gamma}_{\alpha, 1 - \mu}(\lambda; t)$. Both of these integral kernels can be called the fading memory functions and are the Stieltjes functions. It is also shown that their Stieltjes character is enough to ensure the non-negativity of the mean square values and higher even moments. The odd moments vanish. Thus, the solution of generalized Fokker-Planck equations can be called the probability density functions. We introduce also the Volterra-Prabhakar function and its generalization which are involved in the definition of $k_{2}(\alpha, \gamma; \lambda; t)$ and generated by it the probability density function $p_2(x, t)$.
\end{abstract}

\begin{keyword}
Distributed order derivative, distributed order Prabhakar derivative, Volterra-Prabhakar function
\end{keyword}

\allowdisplaybreaks

\maketitle

\section{Introduction}
The anomalous diffusion is characterized by the power-law mean square displacement (MSD), {\it i.e.}, $\langle x^{2}(t) \rangle \simeq t^{\mu}$. It encompasses the sub-diffusion ($\mu\in(0, 1)$) and super-diffusion ($\mu\in(1, 2)$). For $\mu = 1$ we have the normal diffusion and $\mu = 2$ it is the ballistic motion. It is known that the sub-diffusion character is obtained by investigating the anomalous diffusion, for example, either with fractional derivative in the Caputo sense or the Prabhakar derivative \cite{RGarra14, AGiusti19, AGiusti20}. These both fractional derivatives can be presented as $\int_0^t k(t-\xi) \partial_{\xi} f(\xi) \D\xi$ in which $k(t)$ for the Caputo derivative reads
\begin{equation}\label{2/09-15}
K_1(\mu; t) = \frac{t^{-\mu}}{\Gamma(1-\mu)}\qquad \text{and} \qquad \hat{K}_1(\mu; s) = s^{\mu - 1}, \quad \mu\in(0, 1),
\end{equation}    
and for the Prabhakar derivative one has $k(t) = K_2(\alpha, \mu, \gamma; \lambda; t)$,
\begin{equation}\label{2/09-16}
K_2(\alpha, \mu, \gamma; \lambda; t) = e^{-\gamma}_{\alpha, 1-\mu}(\lambda; t)
\qquad \text{and} \qquad \hat{K}_{2}(\alpha, \mu, \gamma; s) = \hat{K}_1(\mu; s) \left(1 + \frac{\lambda}{s^{\alpha}}\right)^{\gamma}, 
\end{equation}
with $\lambda > 0$, $\mu \in(0, 1)$, and $\alpha \in(0, 1)$. The Prabhakar function $e_{\alpha, \beta}^{\gamma}(t;\lambda)$ is equal to
\begin{equation}\label{eq:1092022-w11}
    e_{\alpha, \beta}^{\gamma}(\lambda; t) = t^{\beta - 1} E_{\alpha, \beta}^{\gamma}(-\lambda t^{\alpha}), \quad \text{where} \quad E_{\alpha, \beta}^{\gamma}(z) = \sum_{r=0}^{\infty} \frac{(\gamma)_r z^{r}}{r! \Gamma(\beta + \alpha r)}
\end{equation}
is the three parameters Mittag-Leffler functions and $(\gamma)_r = \Gamma(\gamma + r)/\Gamma(\gamma)$ is the Pochhammer (raising) symbol. The symbol '$\wedge$' denotes the Laplace transform of $f(t)$, this is $\hat{f}(s) = \mathscr{L}[f(t); s]$, where the Laplace pairs are $s\div t$. From Eqs. \eqref{2/09-16} and \eqref{eq:1092022-w11} it follows that $K_2(\alpha, \mu, \gamma; \lambda; t)$ goes to $K_1(\mu; t)$ either for $\lambda = 0$ or $t\ll 1$ because $E_{\alpha, 1-\mu}^{\gamma}(-\lambda t^{\alpha}) \simeq 1/\Gamma(1-\mu)$. For more information about the Mittag-Leffler function we refer to \cite{RGorenflo20, HJHaubold11, KGorska20a, KGorska19, KGorska21}. 

Despite the power-law MSD, in many experiments it is also observed a ultraslow-diffusion, for which MSD grows logarithmically with time, $\langle x^{2}(t)\rangle \simeq \ln^{\nu}(t)$. For instance, it is observed in the Sinai model \cite{YGSinai82} describing the one-dimensional thermal random motion of a particle in a random potential, which is also related to the random-field Ising models \cite{Fisher_PRE} and mechanical DNA unzipping \cite{Kafri_PRL,Walter_PRE}, in the motion in aging environments \cite{Laloux_PRE,Lomholt_PRL}, in iterated maps \cite{JDraeger00}, as well as in annealed (renewal) continuous time random walks with logarithmic waiting time distribution \cite{Godec_JPA}, to name a few. It is generated by the distributed order memory kernels introduced in Refs. \cite{AVChechkin02, AVChechkin03} and, next, considered in, e.g., \cite{TSandev15, TSandev17}, in which the authors took the integral over $\mu\in(0, 1)$ of Eq. \eqref{2/09-15},
\begin{equation}\label{3/09-2}
k_{1}(t) = \int_{0}^{1} \frac{t^{-\mu}}{\Gamma(1- \mu)} \D\mu \qquad \text{and} \qquad \hat{k}_{1}(s) = \int_{0}^{1} s^{\mu - 1}\D\mu = \frac{s-1}{s \ln s}.
\end{equation} 
We can also integrate Eq. \eqref{2/09-16} over $\mu\in(0, 1)$ and consider the so-called distributed order Prabhakar derivative:
\begin{equation}\label{29/08-15}
k_{2}(\alpha, \gamma; \lambda; t) = \int_{0}^{1} e^{-\gamma}_{\alpha, 1 - \mu}(\lambda; t) \D\mu 
\qquad \text{and} \qquad \hat{k}_{2}(\alpha, \gamma; \lambda; s) = \hat{k}_{1}(s) \left(1 + \frac{\lambda}{s^{\alpha}}\right)^{\gamma},
\end{equation}
where $\alpha, \gamma\in(0, 1)$ and $\lambda > 0$. Notice that the integral in Eqs. \eqref{3/09-2} and \eqref{29/08-15}  can be arbitrarily changed by setting $\beta = \mu + p$, $p \in\mathbb{N}$, which is used to determine the higher types of fractional derivatives. For instance, $p=0$ was used in the anomalous diffusion with distributed order derivative \cite{AVChechkin02, AVChechkin03, TSandev17} or in its Langevin pictures \cite{TSandev14}. The special case with $p = 1$ is used in \cite{TMAtanackovic09a, TMAtanackovic09b, TSandev19}.

The memory kernel is involved in the integro-differential equation being the Volterra type,
\begin{equation}\label{2/09-18}
\int_{0}^{t} k(t-\xi) \partial_{\xi} p(x, \xi) \D\xi = B \partial_{x}^{2} p(x, t),
\end{equation}
with $B$ denoting the diffusion coefficient which is a positive constant. Its fundamental solution in the Fourier-Laplace space reads
\begin{equation}\label{eq:20221006-w1a}
\tilde{\hat{p}}(\kappa, s) = \frac{\hat{k}(s)}{s \hat{k}(s) + B \kappa^{2}}
\end{equation}
and taking its inverse Fourier transform we get
\begin{align}\label{29/08-12}
\hat{p}(x, s) = \frac{1}{2 s} \sqrt{\frac{s \hat{k}(s)}{B}}\, \E^{- |x| \sqrt{\frac{s \hat{k}(s)}{B}}}.
\end{align}
Due to the Bernstein theorem (\ref{app1}) we conclude that $p(x, t)$ is a probability density function (PDF) if $\hat{p}(x, s)$ is a completely monotonic function (CMF), {\it i.e.}, it is a non-negative function belonging to $C^{\infty}$ whose all derivative alternate. It is satisfied in twofold: either (a) for $[\hat{k}(s)/s]^{1/2}$ being CMF and $[s \hat{k}(s)]^{1/2}$ being the Bernstein function (BF) or (b) for $[s \hat{k}(s)]^{1/2}$ being the completely Bernstein function (CBF) \cite{KGorska20}. 
\begin{remark}\label{rem1-16Dec}
{\rm 
From (b) and the property (a2) in \ref{app1} follows that $s\hat{k}(s) = \hat{\Psi}(s)$ is CBF, which implies that there exists conjugated with it a CBF function $\hat{\Phi}(s) = s/\hat{\Psi}(s) = 1/\hat{k}(s)$ \cite[Proposition 7.1]{RLSchilling12}. Moreover, from \cite[Theory 7.3]{RLSchilling12} it appears that $\hat{k}(s)$ is a Stieltjes function (SF) which with conjugated with it another SF $\hat{M}(s) = 1/\hat{\Phi}(s)$ forms the Sonnine pair \cite{AHanyga20, AHanyga21} satisfying the Sonnine equation
\begin{equation}\label{3/09-1}
\hat{k}(s)\hat{M}(s) = s^{-1}. 
\end{equation}
Thus, Eq. \eqref{29/08-12} solves also 
\begin{equation}\label{2/09-18a}
p(x, t) = p_0(x) + \int_{0}^{t} M(t -\xi) B \partial_{x}^{2} p(x, \xi) \D\xi, 
\end{equation}
where $p_0(x) = p(x, 0)$ for the fundamental solution is the $\delta$-Dirac distribution. Notice that Eq. \eqref{2/09-18a} is the Sonnine partner of Eq. \eqref{2/09-18} and can be obtained by inserting Eq. \eqref{3/09-1} into Eq. \eqref{2/09-18}, as is shown in \cite{KGorska20a}.
}
\end{remark}
The brief description of CMF, BF, CBF, and SF are presented in \ref{app1}. 

\begin{remark}\nonumber%\label{rem2-16Dec}
{\rm 
We note that the generalized diffusion equation (\ref{2/09-18}) can be obtained within the continuous time random walk theory, by parametrizing the random walk $x(t)$ in terms of the number of steps $u$ via the following coupled Langevin equations \cite{Fogedby_PRE}, see also \cite{TS_ZT_book},
\begin{align*}%\label{LEs}
    \left\lbrace\begin{array}{ll}
         \dot{x}(u)=\zeta(u), \\
         \dot{t}(u)=\xi(u).
    \end{array}\right.
\end{align*}
Here $u$ is called the operational time, which is connected to the physical time by the total $t(u)=\int_{0}^{u}\xi(u')\,du'$ of the individual waiting times $\xi$ for each step. Furthermore, $\zeta(u)$ is a white Gaussian noise ($\langle\zeta(u)\rangle=0$ and $\langle\zeta(u)\zeta(u')\rangle = 2\delta(u-u')$), while $\xi(u)$ is a generalized stable L\'evy noise with a characteristic function $\hat{L}(u,s) = \exp[-us\hat{k}(s)]$ and L\'evy exponent $\hat{\Psi}(s)=s\hat{k}(s)$. This so-called subordination approach can also be used to show the non-negativity of the solution.

\noindent
Therefore, for the memory kernels (\ref{3/09-2}) and (\ref{29/08-15}) the corresponding L\'evy exponents read as $\hat{\Psi}_1(s) = (s-1)/\ln{s}$ and $\hat{\Psi}_2(\alpha, \gamma; \lambda; s) = \hat{\Psi}_1(s)\,(1+ \lambda s^{-\alpha})^{\gamma}$, respectively.}
\end{remark}

In this paper, we give the exact solution of the generalized Fokker-Planck equation with distributed order derivatives as well as calculate the exact form of its MSD. The paper is organized as follows. Sec. \ref{sec2} is the mathematical background of further consideration. We introduce the Volterra-Prabhakar function and its generalization as well as we study their properties. In Sec. \ref{sec3} it is shown that the Volterra-Prabhakar functions appear in the definition of memory kernels $k_1(t)$, $k_2(\alpha, \gamma; \lambda; t)$ and their Sonnie partners. The MSDs as well as the higher moments are calculated in Sec. \ref{sec4} whereas the PDFs is found in Sec. \ref{sec5}. The paper is concluded in Sec. \ref{sec6}.

\section{Volterra-Prabhakar function}\label{sec2}

Volterra's function is defined as follows \cite{Erdei}
\begin{equation}\label{eq:31082022-w10}
\mu(t, \beta, \alpha) = \frac{1}{\Gamma(1+\beta)} \int_{0}^{\infty} \frac{t^{u + \alpha}\, u^{\beta}}{\Gamma(u + \alpha + 1)} \D u, \qquad \RE(\beta) > -1 \quad \text{and} \quad t > 0,
\end{equation}
whose particular cases are
\begin{align*}
%\begin{split}\label{eq:31082022-w10a}
\alpha = \beta = 0: \qquad & \nu(t) = \mu(t, 0, 0), \\
\alpha \neq 0, \, \beta = 0: \qquad & \nu(t, \alpha) = \mu(t, 0, \alpha), \\
\alpha = 0, \, \beta\neq 0: \qquad & \mu(t, \beta) = \mu(t, \beta, 0).
%\end{split}
\end{align*}
The Laplace transform of the Volterra's function $\mu(t, \beta, \alpha)$ is given by \cite{Erdei}
\begin{equation}\label{eq:31082022-w11}
\mathscr{L}[\mu(t, \beta, \alpha); s] = \frac{1}{s^{\alpha + 1} \ln^{\beta + 1}s}.
\end{equation}
More information about Volterra's function can be found in Refs. \cite{Erdei,RGarrappa16a,AApelblat10,AApelblat13,KMehrez18}. Here, we present some properties of the Volterra's function, which are used in the paper.

\begin{proposition}\label{23102022-prop1}
{\rm For $t > 0$, $\alpha > 0$, and $p\in\mathbb{R}$ we have
\begin{equation}\label{eq:2092022-w4}
\int_{0}^{t} \xi^{\alpha - 1} \nu(t-\xi, p) \D\xi = \Gamma(\alpha)\, \nu(t, \alpha + p)
\end{equation}
and in particular 
\begin{equation*}%\label{eq:23102022-w1}
\int_{0}^{t} \xi^{\alpha - 1} \nu(t-\xi, -\alpha) \D\xi = \Gamma(\alpha)\, \nu(t).
\end{equation*}
}
\end{proposition}
{\em Proof.} Proposition \ref{23102022-prop1} can be proved by direct calculations in which we involve the definition of the Euler's Beta function $B(x, y) = \Gamma(x) \Gamma(y)/\Gamma(x + y)$. These calculations read
\begin{align}\label{eq:23102022-w2}
\begin{split}
\int_{0}^{t} \xi^{\alpha - 1} \nu(t-\xi, p) \D\xi &= \int_{0}^{\infty} \frac{1}{\Gamma(1 + p + u)}\left[\int_{0}^{t}  \xi^{\alpha - 1} (t-\xi)^{u + p} \D\xi\right] \D u \\
& = \int_{0}^{\infty} \frac{t^{u + \alpha + p}}{\Gamma(1 + p + u)} \, B(\alpha, 1 + p + u)\, \D u \\
& = \Gamma(\alpha)\, \int_{0}^{\infty} \frac{t^{u + \alpha + p}}{\Gamma(1 + \alpha + p + u)}\, \D u. 
\end{split}
\end{align}
Using the definition of Volterra's $\nu$ function for the integral in Eq. \eqref{eq:23102022-w2} we complete the proof. \qed

\begin{proposition}\label{prop2}
{\rm
For $t > 0$, $\alpha\in(0, 1)$, $\gamma\in\mathbb{R}$, and $p\in\mathbb{R}$ we get
\begin{itemize}
    \item [(a)]
\begin{equation}\label{eq:2092022-w2}
\int_{0}^{t} \nu(t-\xi, p)\, e_{\alpha, 0}^{\gamma}(\lambda; \xi) \D\xi
= \sum_{n=0}^{\infty} \frac{(-\lambda)^{n}}{n!} (\gamma)_{n}\, \nu(t, \alpha n + p), 
\end{equation}

\item[(b)]
\begin{equation}\label{eq:31082022-w12}
\sum_{n=0}^{\infty} \frac{(-\lambda)^{n}}{n!} (\gamma)_{n}\, \nu(t, \alpha n + p) = \int_{0}^{\infty} e_{\alpha, u + p + 1}^{\gamma}(t;\lambda) \D u
\end{equation}
where the Pochhammer (raising) symbol $(\gamma)_{n}$ is defined with Eq. \eqref{eq:1092022-w11}.
\end{itemize}}
\end{proposition}
{\em Proof.} To prove Proposition \ref{prop2}(a) we use the series form of the Prabhakar functions \eqref{eq:1092022-w11} and change (in a legitimate way) the order of integral and series. Hence, we get
\begin{equation*}%\label{eq:23102022-w4}
\int_{0}^{t} \nu(t-\xi, p) e_{\alpha, 0}^{\gamma}(\lambda; \xi) \D\xi = \sum_{n=0}^{\infty} \frac{(-\lambda)^{n}}{n!\, \Gamma(\alpha n)} (\gamma)_{n} \int_{0}^{t} \xi^{\alpha n - 1} \nu(t-\xi, p) \D\xi.
\end{equation*}
The use of Eq. \eqref{eq:2092022-w4} allows one to obtain Eq. \eqref{eq:2092022-w2}. Thereafter, representing the right-hand side (rhs) of Eq. \eqref{eq:31082022-w12} by the definition of Volterra's function and changing the order of series and integral we have
\begin{equation*}
    \sum_{n=0}^{\infty} \frac{(-\lambda)^{n}}{n!} (\gamma)_{n}\, \nu(t, \alpha n + p) = \int_{0}^{\infty} t^{u+p} \sum_{n=0}^{\infty} \frac{(\gamma)_n (-\lambda t^{\alpha})^n}{n!\,\Gamma(u + p+ \alpha n + 1)} \D u,
\end{equation*}
where the series constitutes the three-parameter Mittag-Leffler function \eqref{eq:1092022-w11}. In this way we proved the Proposition \ref{prop2}(b).  \qed

Let the function 
\begin{equation}\label{eq:12092022-w1}
\epsilon_{\alpha,\, p}^{\gamma}(\lambda; t) = \int_{0}^{\infty} e_{\alpha, u + p + 1}^{\gamma}(\lambda; t) \D u
\end{equation}
be called the Volterra-Prabhakar function which for $\gamma = 0$ and/or $\lambda = 0$ reduces to $\nu (t,p)$. It has the following properties:\\
\noindent
{\bf (A)} Applying the Leibnitz rule for differentiation under the integral sign in \eqref{eq:12092022-w1}, this is  $\frac{\D^{\,n}}{\D t^n} e_{\alpha, \beta}^{\gamma}(\lambda; t) = e_{\alpha, \beta-n}^{\gamma}(\lambda; t)$ \cite{RGorenflo20, HJHaubold11}, we obtain the following recurrence relation:
\begin{equation*}
\frac{\D^{\, n}}{\D t^n} [\epsilon_{\alpha,\, p}^{\gamma}(\lambda; t)]=\epsilon_{\alpha,\, p-n}^{\gamma}(\lambda; t) \quad \text{for} \quad n \in\mathbb{N}_{0}.  
 \end{equation*}
In particular, $\frac{\D^{\, n}}{\D t^n} [\epsilon_{\alpha,\, n}^{\gamma}(\lambda; t)] = \epsilon_{\alpha,\, 0}^{\gamma}(\lambda; t)$ , $\frac{\D^{\, n}}{\D t^n} [\epsilon_{\alpha,\, 2n}^{\gamma}(\lambda; t)] = \epsilon_{\alpha,\, n}^{\gamma}(\lambda; t)$,  $\frac{\D^{\, n}}{\D t^n} [\epsilon_{\alpha,\, 3n}^{\gamma}(\lambda; t)] = \epsilon_{\alpha,\, 2n}^{\gamma}(\lambda; t) $, etc., {\it i.e.}, by induction we obtain the general form 
 \begin{equation*}
    \frac{\D^{\, n}}{\D t^n} [\epsilon_{\alpha,\, kn}^{\gamma}(\lambda; t)] = \epsilon_{\alpha,\, (k-1)n}^{\gamma}(\lambda; t), \quad k \in \mathbb{N}. 
 \end{equation*}

\noindent
{\bf (B)} Next, we give a convolution relation between Volterra, Prabhakar, and Volterra-Prabhakar functions.
 Since,
 \begin{equation*}
\mathscr{L}\left[\epsilon_{\alpha,\, p}^{\gamma}(\lambda; t); s\right]\mathscr{L}\left[\mu(t,\beta-1,\alpha-1);s]=\mathscr{L}\left[e_{\alpha,\, p}^{\gamma}(\lambda; t)\right]; s\right]\mathscr{L}\left[\mu(t,\beta,\alpha);s\right],
\end{equation*}
by the convolution theorem of the Laplace transform, we obtain,
\begin{equation*}
\epsilon_{\alpha,\, p}^{\gamma}(\lambda; t)\star \mu(t,\beta-1,\alpha-1)= e_{\alpha,\, p}^{\gamma}(\lambda; t)\star\mu(t,\beta,\alpha).
 \end{equation*}
 Furthermore, from
 \begin{equation*}
\mathscr{L}\left[\epsilon_{\alpha,\, p}^{\gamma}(\lambda; t); s\right]\mathscr{L}\left[\epsilon_{\alpha,\,p'}^{\gamma'}(\lambda; t); s\right]=\mathscr{L}\left[\epsilon_{\alpha,\, p+p'}^{\gamma+\gamma'}(\lambda; t); s\right]\mathscr{L}\left[\nu(t);s\right],
\end{equation*}
we get the convolution semigroup property
\begin{equation*}
 \epsilon_{\alpha,\, p}^{\gamma}(\lambda; t)\star\epsilon_{\alpha,\, p'}^{\gamma'}(\lambda; t) = \epsilon_{\alpha,\, p+p'}^{\gamma+\gamma'}(\lambda; t)\star\nu(t).   
\end{equation*}
In particular,
\begin{equation*}
\epsilon_{\alpha,\, p}^{\gamma}(\lambda; t)\star\epsilon_{\alpha,\, -p}^{-\gamma}(\lambda; t) = \mu(t,1,1). 
\end{equation*}
 In order to evaluate the integral representation of $\epsilon_{\alpha,\, p}^{\gamma}(\lambda; t)$ for $0<\alpha<1$, we consider the inverse transform of the Volterra-Prabhakar function.
 
 \begin{proposition}\label{prop3a} 
{\rm
The Laplace transform of the Volterra-Prabhakar function $\epsilon_{\alpha,\, p}^{\gamma}(\lambda; t)$ is given by
\begin{equation}\label{13/09-2}
\mathscr{L}\left[\epsilon_{\alpha,\, p}^{\gamma}(\lambda; t); s\right] = \frac{s^{\alpha\gamma-p-1}}{(s^{\alpha}+\lambda)^{\gamma} \ln{s}}= \mathscr{L}\left[e_{\alpha,\, p}^{\gamma}(\lambda; t)\star \nu(t); s\right].
\end{equation}}
\end{proposition}
{\em Proof.} Notice that $\mathscr{L}[\epsilon_{\alpha,\, p}^{\gamma}(\lambda; t); s]$ contains two integrals: one of them is placed in the direct Laplace transform and another one is settled in the definition of  the Volterra-Prabhakar function.
Changing the order of these integrals the proof of Proposition \ref{prop3a} follows immediately. \qed 
\smallskip

Notice that Eq. \eqref{13/09-2} can be also presented as
\begin{equation*}%\label{13/09-3}
    \mathscr{L}[\epsilon_{\alpha,\, p}^{-\gamma}(\lambda; t); s] = s^{-p}(s-1)^{-1} \hat{k}_{2}(\alpha, \gamma; \lambda; s).
\end{equation*}
The use of Eq. \eqref{29/08-15} in which by appropriate chose of the values of $\alpha$, $\gamma$, and $\lambda$, enables us to write
\begin{equation*}%\label{13/09-4}
    \mathscr{L}[\epsilon_{1,\, p}^{-1}(-1; t); s] = s^{-p-1} \hat{k}_{1}(s).
\end{equation*}

By using the already established integral representation results and certain earlier investigations by Stanković, Tomovski et al. \cite{TPS 2014} in a lucid and transparent way, gave an elegant proof, that the function $e_{\alpha, \beta}^{\gamma}(\lambda; t)$ is CMF under the conditions $\alpha, \beta \in (0,1)$, $\gamma>0$ and $\alpha\gamma\leq \beta$. Alternative proof can be obtained by taking the product of two CMFs, {\it i.e.}, $t^{\beta-1}$ and three parameters Mittag-Leffler function $E_{\alpha, \beta}^{\gamma}(-x)$. The mathematical rigorous proof, being the extension on Pollard's proof presented in \cite{HPollard48}, that $E_{\alpha, \beta}^{\gamma}(-x)$ is CMF can be found in \cite{KGorska21}.

 \begin{example}\nonumber%\label{15/09-1ex}
 {\rm
 Using Bromwich integral for inverse Laplace transform of \eqref{eq:12092022-w1}, we will find integral representation for Volterra-Prabhakar function with $\lambda=1$ and $0<\alpha\leq 1$. The complex integral can be evaluated by taking into account that the integrand has the branch point at $s=0$ and the pole at $s=1$. The point $s = \exp(\I\!\pi)=-1$, (for $\alpha = 1$) is isolated singular point. For all non-integer values of $\alpha$, the power $s^\alpha$ is given by $s^\alpha=\lvert s \rvert ^\alpha e^{\I\alpha\arg(s)}$, where $\lvert \arg(s)\rvert <\pi$, that is, in the complex $s$-plane cut along the negative real axis. The Bromwich contour used in the integration is presented in Figure \ref{fig1}. 
 \begin{figure}%[!h]
     \centering
     \includegraphics[scale=0.6]{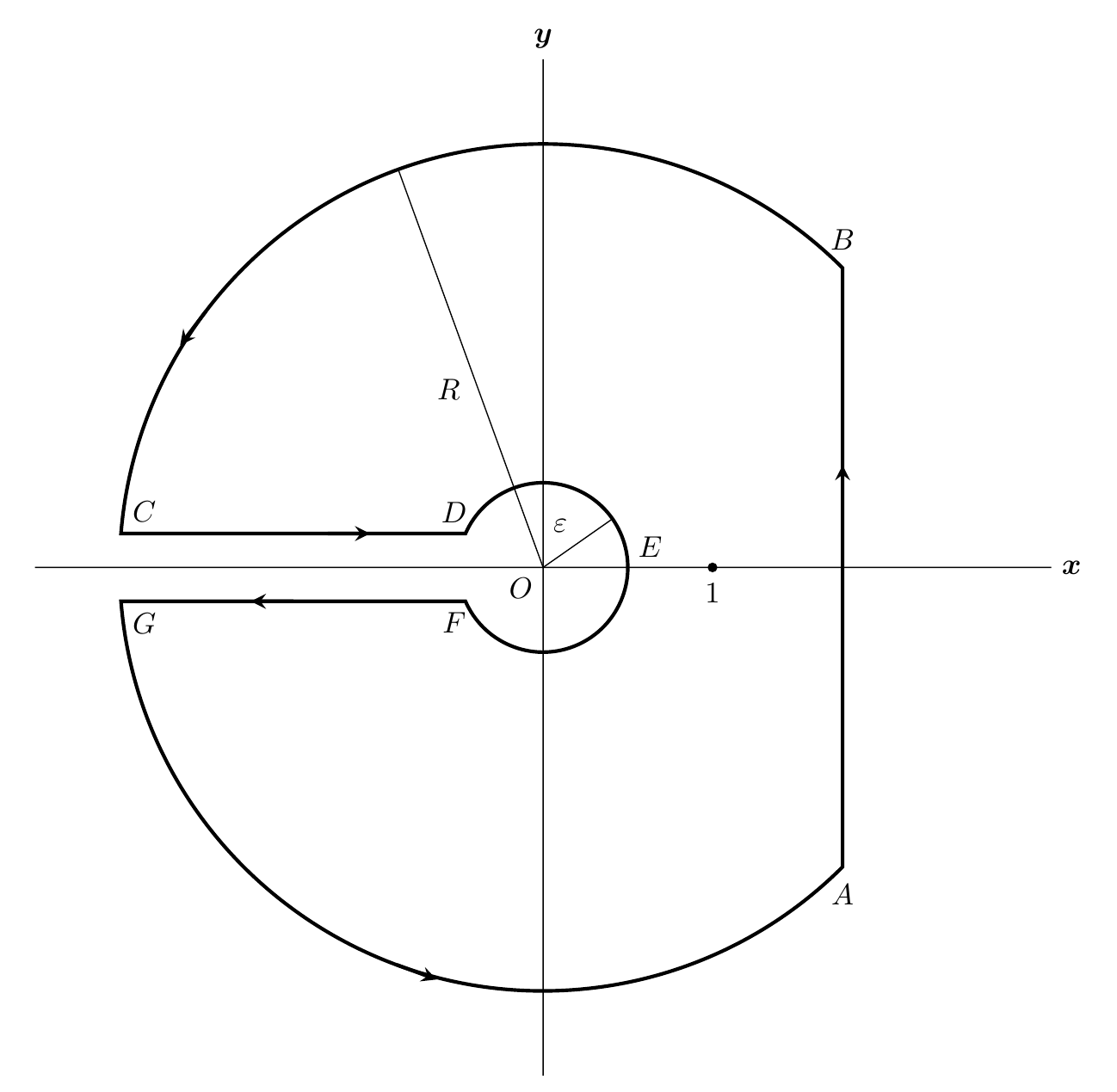}
     \caption{The Bromwich contour}
     \label{fig1}
 \end{figure}
 The contour consists of the straight line $AB$, a large semi-circle of radius $R$, a small circle about the origin, and a cut along the negative axis. According to the Jordan lemma, the integrals over the large semi-circle parts $BC$ and $GA$ vanish as $R\to\infty$. Thus contributions to the integral come only from the loop $H$, starting from $CD$, encircles the circular disk $DEF$ and finishes to the lower side of the negative real half-axis, {\it i.e.}, $FG$ parts of the contour and from the pole at $s = 1$. The residue at this point is $\E^t/2^{\gamma}$.

Therefore
 \begin{equation*}%\label{18/09-2}
\epsilon_{\alpha,\, p}^{\gamma}(1; t) \equiv \epsilon_{\alpha,\, p}^{\gamma}(t) = \frac{1}{2\pi\!\I}\int_{c-\I\!\infty}^{c+\I\!\infty}\E^{st}\frac{s^{\alpha\gamma-p}}{(s^{\alpha}+1)^{\gamma} s\ln{s}}\D s = \frac{\E^t}{2^\gamma}+f_{\alpha,p}^\gamma (t), 
 \end{equation*}
 where
 \begin{equation*}
f_{\alpha,p}^\gamma(t)=\frac{1}{2\pi\!\I}\int_{H}\E^{st}\frac{s^{\alpha\gamma-p}}{(s^{\alpha}+1)^{\gamma} s\ln{s}} \D s = \mathscr{L}\left[K_{\alpha,\, p}^{\gamma}(r); t\right].
 \end{equation*}
By applying the Titchmarsh formula \cite[Eq.(11.6.5) on p. 316]{Ttc37}, it follows
\begin{align}\label{18/09-1}
    \begin{split}
K_{\alpha, p}^{\gamma}(r)&=-\frac{1}{\pi}\IM\left\{\frac{r^{\alpha\gamma-p}\E^{\I\!\pi(\alpha\gamma-p)}}{(r^{\alpha}\E^{\I\!\pi\alpha}+1)^\gamma(r\E^{\I\!\pi})\ln(r\E^{\I\!\pi})}\right\} \\
& =-\frac{r^{\alpha\gamma-p-1}}{\pi}\IM\left\{\frac{\E^{\I\!\pi(\alpha\gamma-p-1)}}{(r^{\alpha} \E^{\I\!\pi\alpha}+1)^\gamma (\I\pi + \ln{r})}\right\}.
\end{split}
 \end{align}
Next, we remove the imaginary part from the denominator. To realize this purpose we multiply and divide Eq. \eqref{18/09-1} by $[r^{\alpha} \exp(-\I\!\pi\alpha) +1]^{\gamma} = |z|^{\gamma} \exp[-\I\!\gamma\theta_{\alpha}(r)]$ and $(-\I\!\pi + \ln r)$. Thus, we get
\begin{equation*}%\label{19/09-1}
K_{\alpha, p}^{\gamma}(r) = - \frac{r^{\alpha\gamma-p-1}}{\pi} \IM\left\{\frac{\E^{\I\!\pi(\alpha\gamma - p -1 )}|z|^{\gamma} \E^{-\I\!\gamma\theta_{\alpha}(r)}(-\I\!\pi + \ln r)}{|z|^{2\gamma} (\pi^2 + \ln^2 r)} \right\}
\end{equation*}
with 
\begin{equation*}%\label{19/09-2}
    |z| = [r^{2\alpha} + 2 r^{\alpha}\cos(\pi\alpha) + 1]^{1/2} \qquad \text{and} \qquad
    \theta_{\alpha}(r) = \arctan\left[\frac{\sin(\pi\alpha)}{\cos(\pi\alpha) + r^{-\alpha}}\right].
\end{equation*}
Then, the spectral function kernel $K_{\alpha, p}^{\gamma}(r)$ can be written as
\begin{equation*}
K_{\alpha, p}^{\gamma}(r) = \frac{r^{\alpha\gamma - p -1}}{\pi} \frac{(\ln r)\sin[\pi(\alpha\gamma - p) - \gamma\theta_{\alpha}(r)] - \pi\cos[\pi(\alpha\gamma - p) - \gamma\theta_{\alpha}(r)]}{[r^{2\alpha} + 2r^{\alpha}\cos(\pi\alpha) + 1]^{\gamma/2} (\pi^2 + \ln^2 r)}.
\end{equation*}
Finally,
\begin{equation}\label{19/09-4}
\epsilon_{\alpha,\, p}^{\gamma}(t) = \frac{\E^t}{2^\gamma}-\int_{0}^{\infty}\E^{-rt}\Tilde{K}_{\alpha,p}^{\gamma}(r)\D r, 
\end{equation}
where $\widetilde{K}_{\alpha,p}^{\gamma}(r) = - {K}_{\alpha,p}^{\gamma}(r).$ Since $\epsilon_{\alpha,\, p}^{\gamma}(0)=0$ then from Eq. \eqref{19/09-4} we obtain the following result
\begin{equation*}%\label{19/09-5}
 \int_{0}^{\infty}\widetilde{K}_{\alpha,p}^{\gamma}(r)\D r=\frac{1}{2^\gamma}.   
\end{equation*}
We ask now, under which conditions on parameters, the kernel $K_{\alpha, p}^{\gamma}(r)$ is a negative function, {\it i.e.}, $\widetilde{K}_{\alpha,p}^{\gamma}(r)$ is positive in respect to $r$? Since $r^{2\alpha} + 2r^{\alpha}\cos(\pi\alpha) + 1\geq r^{2\alpha}-2r^{\alpha}+1=(r^{\alpha}-1)^2>0$ and $\pi^2 + \ln^2 r>0$, one option $K_{\alpha, p}^{\gamma}(r)$  to be negative is $\alpha\in (0,1/2]$, $\gamma>0$, and $\alpha\gamma=p$ or $\alpha\gamma-p=2k$, where $k$ is integer number. In this case, $2^{\gamma}\widetilde{K}_{\alpha,p}^{\gamma}(r)=2^{\gamma}\widetilde{K}_{\alpha,\alpha\gamma}^{\gamma}(r)>0$ or $2^{\gamma}\widetilde{K}_{\alpha,p}^{\gamma}(r)=2^{\gamma}\widetilde{K}_{\alpha,\alpha\gamma-k}^{\gamma}(r)>0$ are the densities of a probability measure concentrated on the positive real line (see Figs \ref{fig2}, \ref{fig3}, and \ref{fig4}). If $p=0$, $\gamma>0$ then we will analyse the sign of the argument $\gamma [\pi\alpha - \theta_{\alpha}(r)]$ of $\sin$ and $\cos$ functions. So, if $\alpha \in (1/2,1)$ and $0<r<1$ then
\begin{equation*}
\tan(\pi\alpha)-\frac{\sin(\pi\alpha)}{\cos(\pi\alpha) + r^{-\alpha}}=\frac{r^{-\alpha}\tan(\pi\alpha)}{\cos(\pi\alpha)+r^{-\alpha}}<0,
\end{equation*}
or $\tan(\pi\alpha)<\frac{\sin(\pi\alpha)}{\cos(\pi\alpha) + r^{-\alpha}}$, {\it i.e.}, $\pi\alpha < \theta_{\alpha}(r)$. Hence, $\gamma [\pi\alpha - \theta_{\alpha}(r)]<0$, {\it i.e.}, $\widetilde{K}_{\alpha,0}^{\gamma}(r)>0$.
So, $2^\gamma\widetilde{K}_{\alpha,0}^{\gamma}(r)$ represents density function for $0<r<1$.}
\end{example}

\begin{figure}[h!]
     \centering
     \begin{subfigure}[b]{0.462\textwidth}
         \centering
         \includegraphics[width=\textwidth]{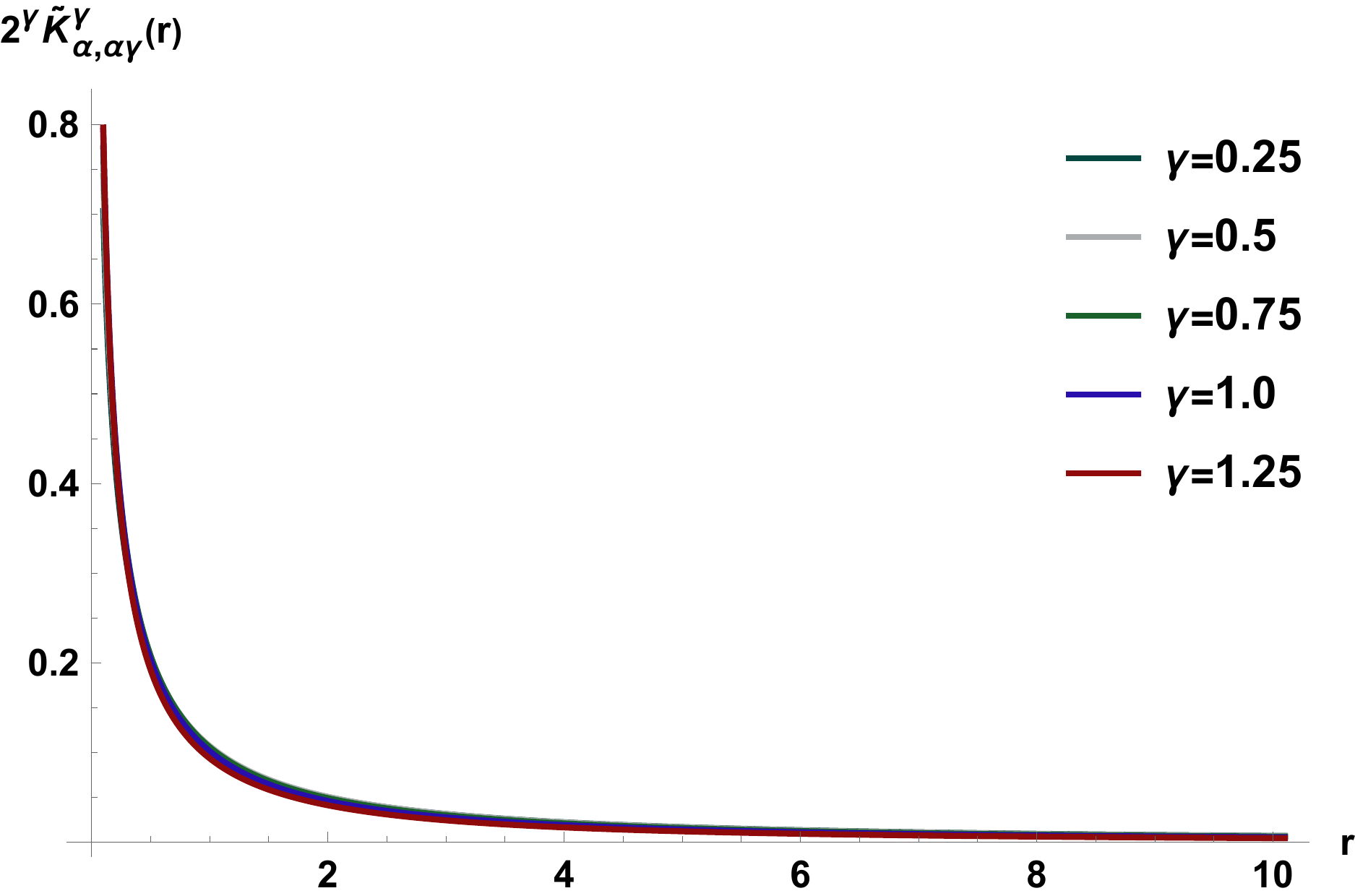}
     \end{subfigure}
     \hfill
     \begin{subfigure}[b]{0.462\textwidth}
         \centering
         \includegraphics[width=\textwidth]{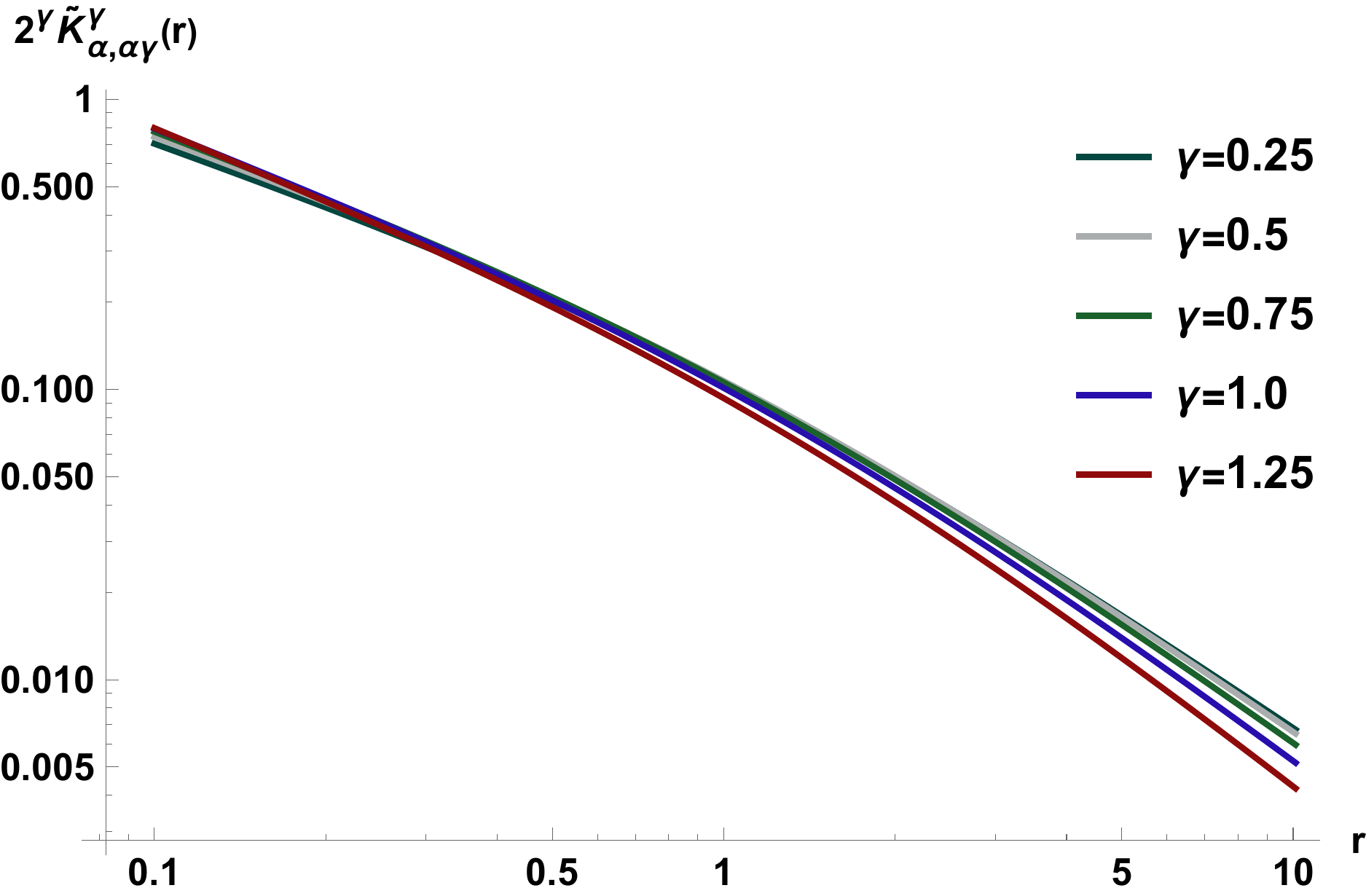}         
     \end{subfigure}
        \caption{\label{fig2} Linear-linear (left) and log-log plot (right) of the function $2^{\gamma}\widetilde{K}_{\alpha,\alpha\gamma}^{\gamma}$ for $\alpha = 0.4$.}        
\end{figure}
\begin{figure}[h!]
     \centering
     \begin{subfigure}[b]{0.462\textwidth}
         \centering
         \includegraphics[width=\textwidth]{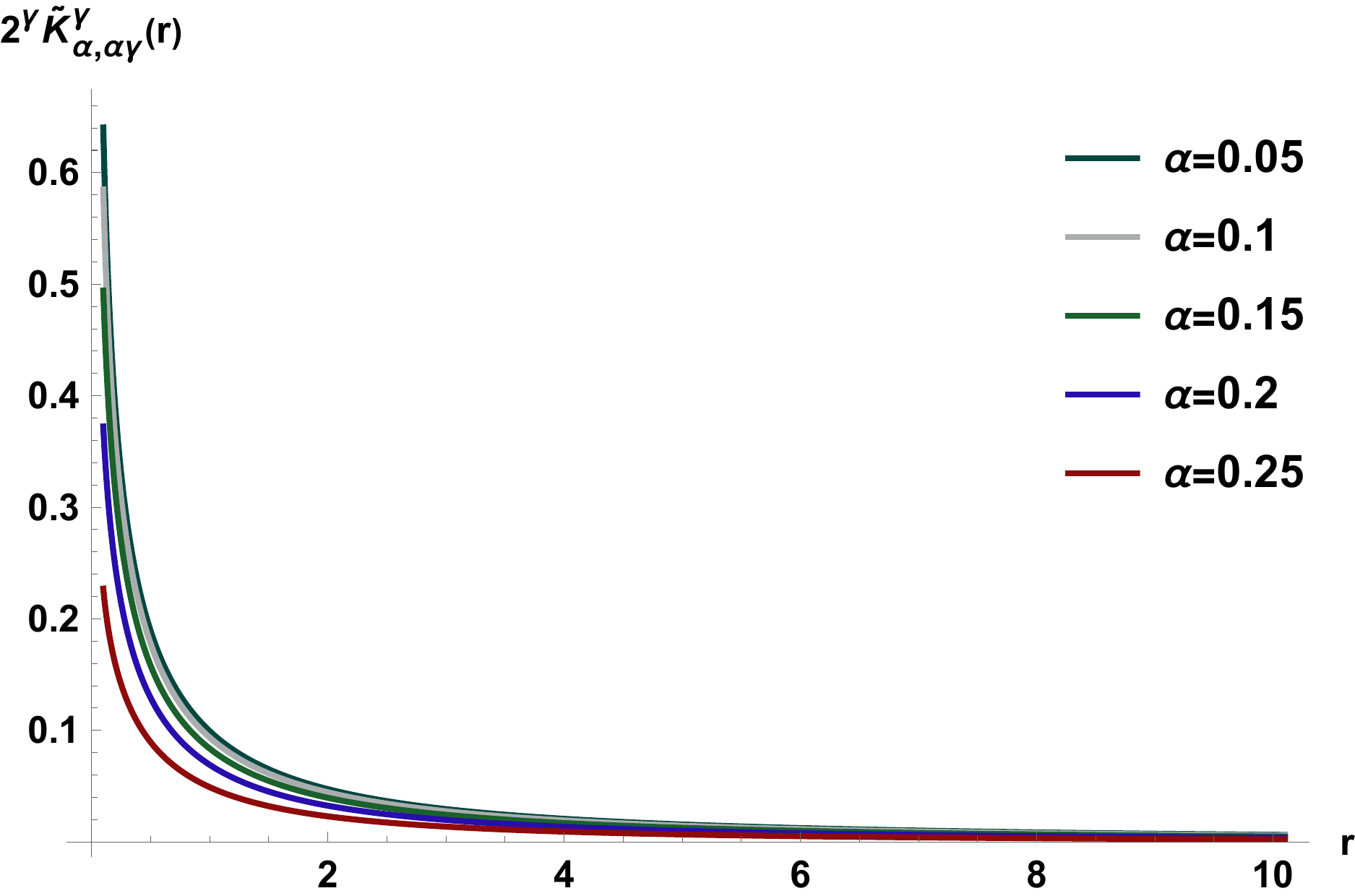}
     \end{subfigure}
     \hfill
     \begin{subfigure}[b]{0.462\textwidth}
         \centering
         \includegraphics[width=\textwidth]{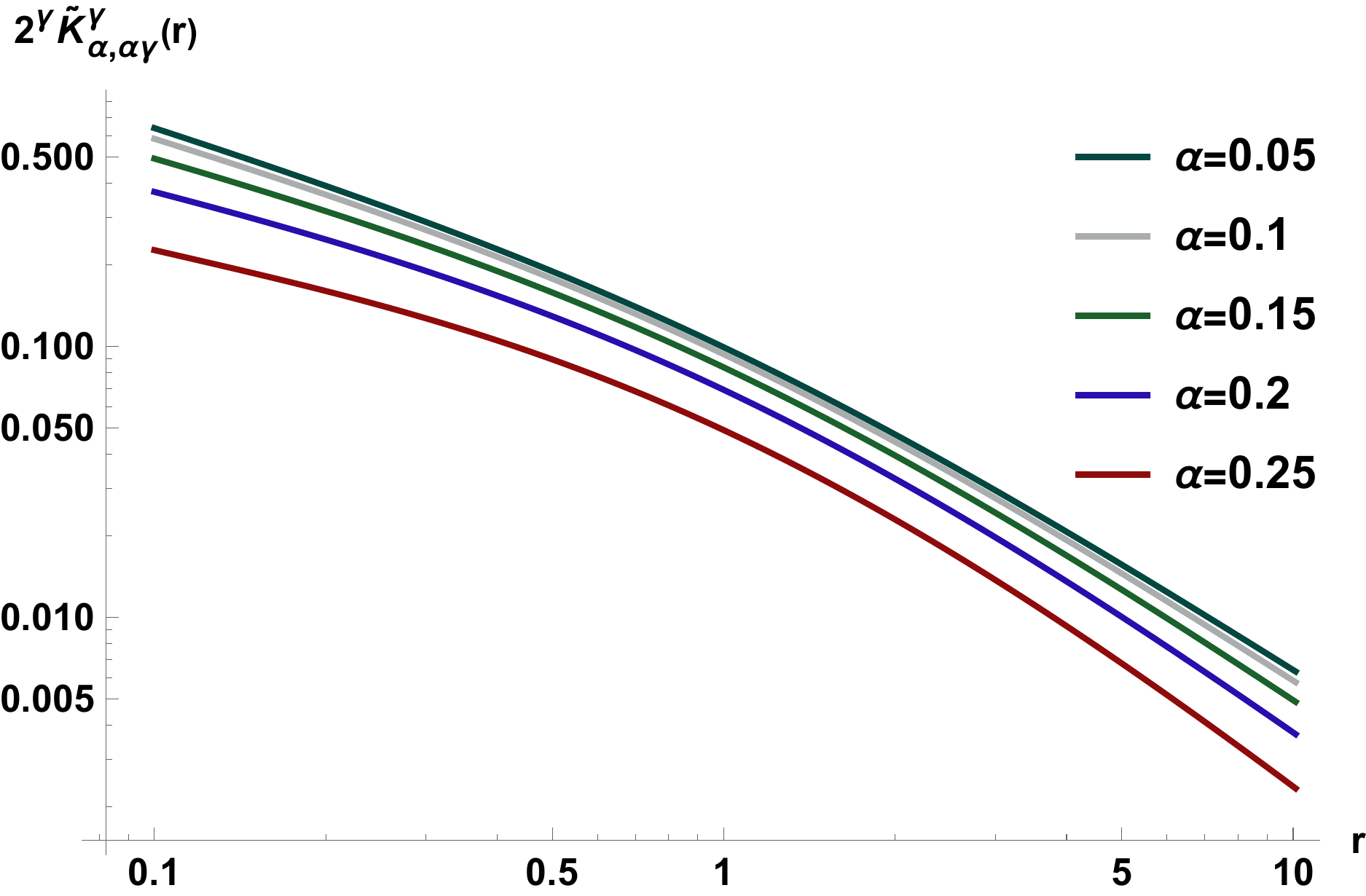}         
     \end{subfigure}
        \caption{\label{fig3} Linear-linear (left) and log-log plot (right) of the function $2^{\gamma}\widetilde{K}_{\alpha,\alpha\gamma}^{\gamma}$ for $\gamma = 3.0$.}        
\end{figure}
\begin{figure}[h!]
     \centering
     \begin{subfigure}[b]{0.462\textwidth}
         \centering
         \includegraphics[width=\textwidth]{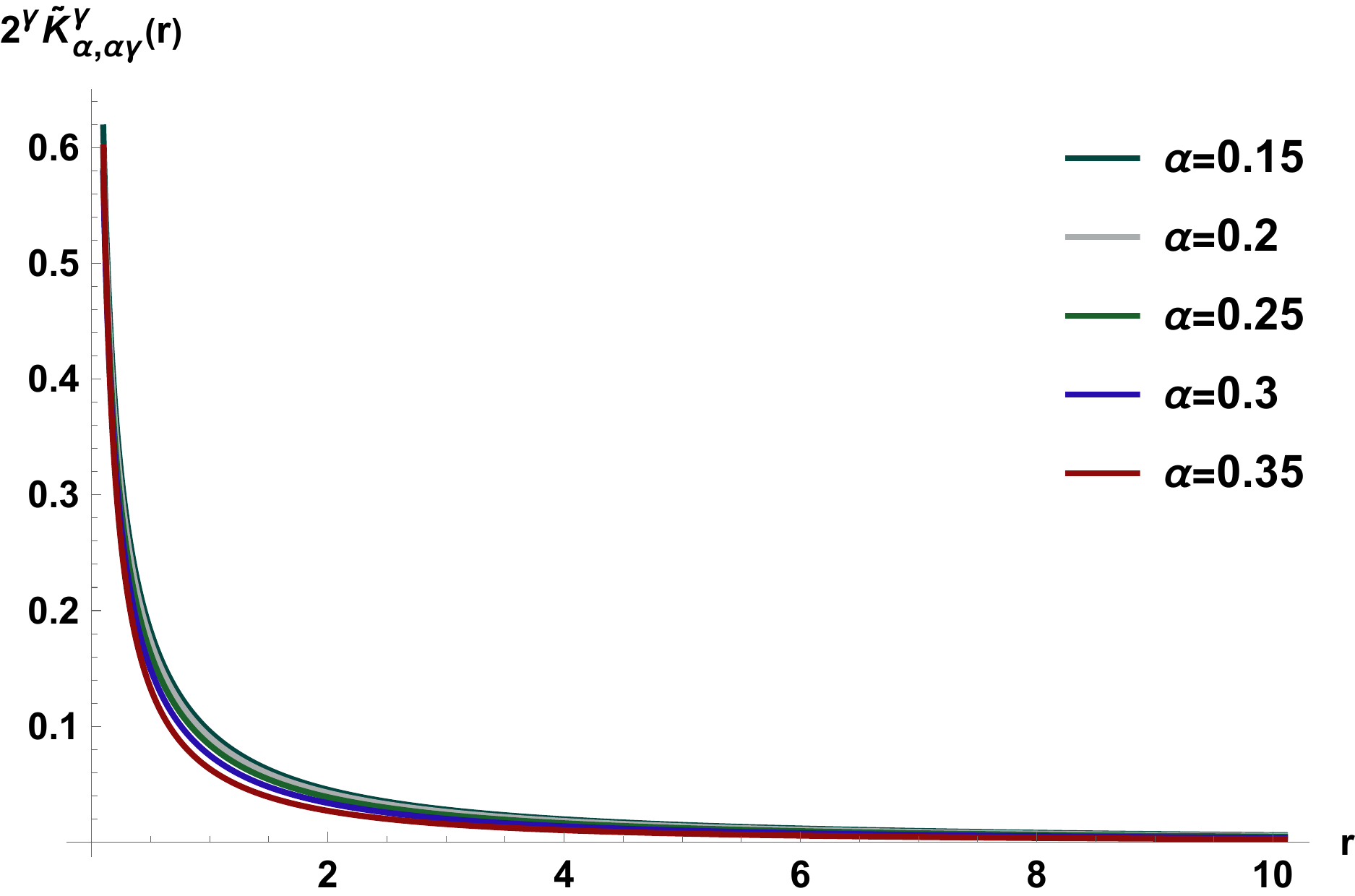}
     \end{subfigure}
     \hfill
     \begin{subfigure}[b]{0.462\textwidth}
         \centering
         \includegraphics[width=\textwidth]{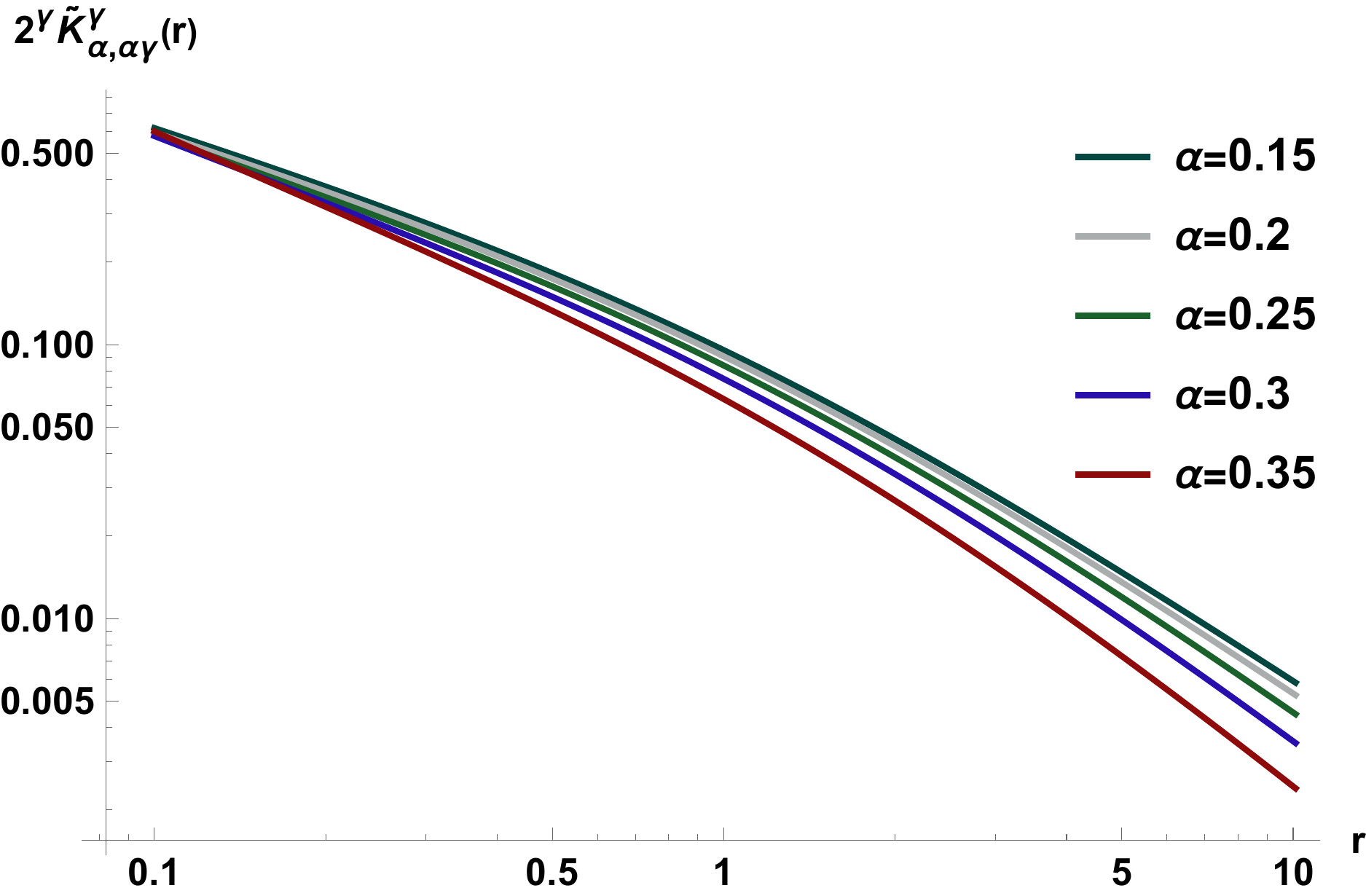}         
     \end{subfigure}
        \caption{\label{fig4} Linear-linear (left) and log-log plot (right) of the function $2^{\gamma}\widetilde{K}_{\alpha,\alpha\gamma}^{\gamma}$ for $\gamma = 2.0$.}        
\end{figure}
\newpage
\noindent
By applying the Bernstein theorem, we obtain the following results:
\begin{proposition}\nonumber%\label{prop4} 
{\rm
Let $\gamma>0$. The following assertions hold true:
\begin{itemize}
    \item [(a)]
 $\epsilon_{\alpha, p}^{\gamma}(t)-\frac{e^{t}}{2^{\gamma}}$ is CM on $(0,\infty)$ for $\alpha\in (0,1/2]$ and $\alpha\gamma-p=2k$, $k\in \mathbb{Z}$;
 \item [(b)]
 $\epsilon_{\alpha,\alpha\gamma}^{\gamma}(t)-\frac{e^{t}}{2^{\gamma}}$ is CM on $(0,\infty)$ for $\alpha\in (0,1/2]$; 
 \item [(c)]
$\epsilon_{\alpha,0}^{\gamma}(t)-\frac{e^{t}}{2^{\gamma}}$ is CM on $(0,1)$ for $\alpha\in (1/2,1]$.
\end{itemize}}
\end{proposition}
Consequently, under the same conditions, the functions of the previous proposition are log–convex, since every CM function is log–convex, see \cite{Wid 41}. In particular, for $\gamma=0$ we obtain the integral
\begin{equation*}%\label{19/09-3}
   \nu(t,p) = \epsilon_{\alpha, p}^{0}(t) = e^{t} - \frac{1}{\pi} \int_{0}^{\infty} \frac{\E^{-rt}}{r^{p+1}} \frac{(\ln r)\sin(\pi p) + \pi \cos(\pi p)}{\pi^2 + \ln^2 r} \D r.
\end{equation*}
Hence, 
\begin{equation*}
  \int_{0}^{\infty}\frac{1}{\pi}r^{p+1}\frac{(\ln r)\sin(\pi p) + \pi \cos(\pi p)}{\pi^2 + \ln^2 r} \D r = 1.  
\end{equation*}
For $p\in[0,1/2]$ the integrand of the last integral is positive, so it represents a PDF. Furthermore, for $\gamma=p=0$, we obtain the Ramanujan integral (see \cite[Eq. (16.2.3) on p. 27]{AApelblat13})
\begin{equation*}
\nu(t)=\epsilon_{\alpha, 0}^{0}(t) = \E^t - \int_{0}^{\infty} \frac{e^{-rt}}{r} \frac{1}{\pi^2 + \ln^2 r} \D r.    
\end{equation*}
Hence,
\begin{equation*}
\int_{0}^{\infty} \frac{1}{r} \frac{1}{\pi^2 + \ln^2 r} \D r = 1.
\end{equation*}
The integrand of the last integral is a positive function, so it represents a PDF.

\begin{proposition} {\rm The following assertions hold true:
\begin{itemize}
    \item [(a)]
  The function $\E^{t}-\nu(t,p)$ is CM and log-convex on $(0,\infty)$, for $p\in [0,1/2]$;
  
    \item [(b)]
  The function $\E^{t}-\nu(t)$ is CM and log-convex on $(0,\infty)$.
 \end{itemize}}
\end{proposition}
The cases $1<\alpha\leq2$ and $\gamma\in \mathbb{N}$ we leave for the reader. For an idea, we refer to Ref. \cite{TPS 2014}.

A more general form of the Volterra-Prabhakar function, associated by \eqref{eq:31082022-w10} is
\begin{equation}
\epsilon_{\alpha,\,\beta,\, p}^{\gamma}(\lambda; t) = %\epsilon(t; \lambda; \alpha, p, \gamma)= 
\int_{0}^{\infty} u^{\beta} e_{\alpha, u + p + 1}^{\gamma}(\lambda; t) \D u.
\label{eq:20221006-w1}
\end{equation}
In particular, $\epsilon_{\alpha,\,0,\, p}^{\gamma}(\lambda; t)=\epsilon_{\alpha,\, p}^{\gamma}(\lambda; t)$. %and $\epsilon_{\alpha,\, \beta,\,p}^{\gamma}(1; t)=\epsilon_{\alpha,\, \beta,\,p}^{\gamma}(t).$

\smallskip
\noindent
Without proof, we will present slight generalizations of Propositions 1 and 2.
\begin{proposition}\label{p:20221009-p1}
{\rm For $t > 0$, $\alpha, \beta > 0$, $ \lambda, \gamma \in  \mathbb{R} $ we have
\begin{itemize}
\item [(a)]
\begin{equation*}
\int_{0}^{t} \xi^{\alpha - 1} \mu(t-\xi, \beta, \alpha) \D\xi = \Gamma(\alpha)\, \mu(t, \beta, 2\alpha),
\label{eq:20221009-w1}
\end{equation*}
\item [(b)]
\begin{equation*}
\int_{0}^{t} \mu(t-\xi, \beta, \alpha)\, e_{\alpha, \beta}^{\gamma}(\lambda; \xi) \D\xi
= \sum_{n=0}^{\infty} \frac{(-\lambda)^{n}}{n!} (\gamma)_{n}\, \mu(t, \beta, \alpha n + \alpha + \beta), 
%\label{eq:20221009-w2}
\end{equation*}
\item [(c)]
\begin{equation*}
\epsilon_{\alpha, \, \beta,\, \alpha+\beta}^{\gamma} (\lambda; t)=\sum_{n=0}^{\infty} \frac{(-\lambda)^{n}}{n!} (\gamma)_{n}\, \mu(t, \beta, \alpha n + \alpha+\beta).
%\label{eq:20221009-w3}
\end{equation*}
\end{itemize}}
\end{proposition}

\begin{proposition}\label{p:20221008-p1}
{\rm 
The Laplace transform of $\epsilon_{\alpha,\,\beta,\, p}^{\gamma}(\lambda; t)$ for $\RE \beta > 0$ reads
\begin{equation*}
\mathscr{L}[\epsilon_{\alpha,\,\beta,\, p}^{\gamma}(\lambda; t); s] = \Gamma(1 + \beta) \frac{s^{\alpha\gamma-p-1}}{(s^\alpha + \lambda)^{\gamma}\, (\ln s)^{1+\beta}}.
   % \label{eq:20221008-w1}
\end{equation*}
Furthermore, the following convolution relation holds true,
\begin{equation*}
\epsilon_{\alpha,\,\beta,\, p}^{\gamma}(\lambda; t)= e_{\alpha, p}^{\gamma}(\lambda; t)\star\mu(t,\beta,0).  
%\label{p:20221008-w10}
\end{equation*}}
\end{proposition}
{\em Proof.} The proposition \ref{p:20221008-p1} can be proved by direct calculations in which we use Eq. \eqref{eq:20221006-w1} and change the order of integral with the integral defining the direct Laplace transform. Hence, we have
\begin{align*}
%\begin{split}
\mathscr{L}[\epsilon_{\alpha,\,\beta,\, p}^{\gamma}(\lambda; t); s] & = \int_0^{\infty} u^{\beta} \left[\int_0^{\infty} \E^{-st} e_{\alpha, u + p+ 1}^{\gamma}(\lambda; t)\D t\right] \D u \\
& = \frac{s^{\alpha\gamma-p-1}}{(s^{\alpha} + \lambda)^{\gamma}} \int_{0}^{\infty} u^{\beta} s^{-u} \D u.\label{eq:20221008-w2}
  %  \end{split}
\end{align*}
The use of $\int_{0}^{\infty} u^{\beta} s^{-u} \D u = \Gamma(1+\beta) (\ln s)^{-1-\beta}$ finishes the proof. \qed 
\begin{proposition}\nonumber%\label{p:20221008-p9}
{\rm For $\alpha>0, \ 0<\RE\beta <1$, $0 < \RE\gamma\leq 1-\RE\beta,\ p\in \mathbb{R} $ the Melline transform of $\epsilon_{\alpha, \,\beta,\, p}^{\gamma}(t)$ is given by
\begin{multline*}
\mathscr{M}[\epsilon_{\alpha,\,\beta,\, p}^{\gamma}(t); s]  =\frac{-\alpha^{\beta}\beta\pi}{[\sin(\pi\beta)]\,\Gamma(1-s)} \sum_{k=0}^{\infty}\binom{-\gamma}{k}(\gamma+2k)\theta_{k}^{\beta-1} \\
 (-\RE\gamma+\frac{p+s}{\alpha}\leq\theta_{k}\leq \frac{p+s}{\alpha}\ ).
\end{multline*}}
\end{proposition}
{\em Proof.} Using the Meline transform formula for Prabhakar function,
\begin{equation*}
    \mathscr{M}[E_{\alpha,\,\alpha^{'}}^{\gamma}(-t); s] =\frac{1}{\Gamma(\gamma)}\frac{\Gamma(s)\Gamma(\gamma-s)}{\Gamma(\alpha^{'}-\alpha s)} \quad (0<\RE s<\RE \gamma),
\end{equation*}
we have
\begin{align*}
    \mathscr{M}[\epsilon_{\alpha,\,\beta,\, p}^{\gamma}(t); s] &= \int_0^{\infty} u^{\beta} \left[\int_0^{\infty} t^{s-1} e_{\alpha, u + p+ 1}^{\gamma}(t)\D t\right] \D u \\  
& = \frac{1}{\alpha \Gamma(\gamma)\Gamma(1-s)}\int_0^{\infty} u^{\beta} \Gamma\left(\frac{u+p+s}{\alpha}\right)\Gamma\left(\gamma-\frac{u+p+s}{\alpha}\right) \D u \\ 
& = \frac{\alpha^{\beta}}{\Gamma(\gamma)\Gamma(1-s)}\int_0^{\infty} u^{\beta} \Gamma\left(u+\frac{p+s}{\alpha}\right)\Gamma\left(\gamma-u-\frac{p+s}{\alpha}\right) \D u. \\ 
 \end{align*}
Let us use $l=\frac{p+s}{\alpha}$ and apply the definition for the gamma function
 \begin{align*}
J & = \int_{0}^{\infty}u^{\beta} \left[\int_{0}^{\infty} \int_{0}^{\infty} \E^{-(x+y)} x^{u+l-1}y^{\gamma-u-l-1}\D x \D y\right]\D u \\
& = \int_{0}^{\infty}\int_{0}^{\infty} \E^{-(x+y)} x^{l-1} y^{\gamma-l-1} \D x \D y\int_{0}^{\infty}u^{\beta}\left(\frac{x}{y}\right)^u \D u \\ 
& = \Gamma(\beta+1) \int_{0}^{\infty} \int_{0}^{\infty} \E^{-(x+y)} x^{l-1}y^{\gamma-l-1} \left(\ln\frac{y}{x}\right)^{-(\beta+1)} \D x \D y.
 \end{align*}
Recently, Maryam Al-Kandari et al. \cite {KHL19} introduced the following integral operator
 \begin{equation*}
 (T_{\phi}f)(x)=\int_{0}^{\infty}\int_{0}^{\infty} \E^{-(\tau_{1}+\tau_{2})} \tau_{1}^{-\alpha} \tau_{2}^{\beta-1} f\left(\frac{x\tau_{1}^{a}}{\tau_{2}^{b}}\right) \D \tau_{1} \D \tau_{2}.
\end{equation*}
Taking $a=b=1$, $x=1$, $\alpha=1-l$, $\beta=\gamma-l$ and $f(\tau_1/\tau_2)=[\ln(\tau_{2}/\tau_{1})]^{-(\beta+1)}$ in (52), we get
\begin{equation*}
J=\Gamma(\beta+1)(T_{\phi} \ln^{-\beta-1})(1).
\end{equation*}
Substituting $x+y=v$ and $y/x = w$, we obtain $x=v/(1+w)$, $y= vw/(1+w)$ with Jacobi-determinant $\frac{\partial(x,y)}{\partial(v,w)} =v/(1+w)^2$, and 
\begin{align*}
J & =\Gamma(\beta+1)\Gamma(\gamma) \int_{0}^{\infty} w^{\gamma-l-1} (1+w)^{-\gamma}(\ln w)^{-(\beta+1)} \D w \\ 
&  =\Gamma(\beta+1)\Gamma(\gamma)(J_{1}+J_{2}), \\
\end{align*}
where 
\begin{equation*}
J_{1}=\int_{0}^{1} w^{\gamma-l-1} (1+w)^{-\gamma}(\ln w)^{-(\beta+1)} \D w
\end{equation*}
and 
\begin{equation*}
 J_{2}=\int_{1}^{\infty} w^{\gamma-l-1} (1+w)^{-\gamma}(\ln w)^{-(\beta+1)} \D w. 
\end{equation*}
To solve the last integrals, we will apply the integral given in \cite{GR07},
\begin{multline*}
\int_{0}^{1}\left(\ln\frac{1}{x}\right)^{r-1}\frac{x^{p-1}}{(1+x^q)^s}\D x=
\Gamma(r)\sum_{k=0}^{\infty}\binom{-s}{k}\frac{1}{(p+kq)^r} \\
(p \ge 0 ,\  q \ge 0 ,\ r \ge 0,\ 0\leq s\leq r + 2)
\end{multline*}
with $r=-\beta$, $p=l$, $q=1$, $s=\gamma$. Namely,
\begin{equation*}
    J_{2}=\int_{0}^{1}\left(\ln\frac{1}{w}\right)^{-(\beta+1)}\frac{w^{l-1}}{(1+w)^\gamma} \D w \\  =\Gamma(-\beta)\sum_{k=0}^{\infty}\binom{-\gamma}{k}(l+k)^{\beta}.
\end{equation*}
Analogously,
\begin{equation*}
 J_{1}= (-1)^{\beta+1}\Gamma(-\beta)\sum_{k=0}^{\infty}\binom{-\gamma}{k}(\gamma-l+k)^{\beta}.
\end{equation*}
Then, by mean theorem of Lagrange, we have
\begin{align*}
J_{1}+J_{2} & =\Gamma(-\beta)\sum_{k=0}^{\infty}\binom{-\gamma}{k}[(l+k)^{\beta}-(-\gamma+l-k)^{\beta}] \\ 
& = -\Gamma(1-\beta) \sum_{k=0}^{\infty} \binom{-\gamma}{k} (\gamma+2k) \theta_{k}^{\beta-1} \\
&(\RE\beta > 0,\ 0 < \RE\gamma\leq 1-\RE\beta, \ -\RE\gamma+l-k\leq\theta_{k}\leq l+k,\ k=0,1,2,\dots). \qquad\qquad\quad \Box
\end{align*}
 
\section{The distributed order Prabhakar kernel and its Sonnine partner}\label{sec3}

The distributed order kernel $\hat{k}_1(s)$ and distributed order Prabhakar kernel $\hat{k}_2(\alpha, \gamma; \lambda; s)$ fulfill Kochubei's limits given by \cite[Eqs. (3.1) and (3.2)]{ANKochubei11} from which it appears that $[s\hat{k}_1(s)]^{-1}$ and $[s\hat{k}_2(\alpha, \gamma; \lambda; s)]^{-1}$ vanishes at $s \gg 1$. Hence, both of them can be called the fading memory function \cite{ECapelas18, DZhao19}. 

In Ref. \cite{TSandev19} it is shown that $\hat{k}_1(s)$ and $\hat{k}_2(\alpha, \gamma; \lambda; s)$ are SFs and they can be presented as $s \hat{k}_{1}(s)/s$ and $s \hat{k}_{2}(\alpha, \gamma; \lambda; s)/s$, where $s \hat{k}_{1}(s)$ and $s\hat{k}_{2}(\alpha, \gamma; \lambda; s)$ are CBFs (the property (a6) in \ref{app1}). It guarantees that Remark \ref{rem1-16Dec} is satisfied which means that exist the Sonnine partner of $\hat{k}_1(s)$ and $\hat{k}_2(\alpha, \gamma; \lambda; s)$. Namely
\begin{equation}\label{1/09-1}
\hat{M}_{1}(s) = \frac{1}{s\hat{k}_1(s)} = \frac{\ln s}{s-1} \quad \text{and} \quad \hat{M}_{2}(\alpha, \gamma; \lambda; s) = \frac{1}{s\hat{k}_2(\alpha, \gamma; \lambda; s)} = \hat{M}_{1}(s) \left(1 + \frac{\lambda}{s^{\alpha}}\right)^{-\gamma},
\end{equation}
and $\alpha, \gamma\in(0, 1)$ with $\lambda > 0$.

The asymptotic behavior of $k_{1}(t)$, $k_{2}(\alpha, \gamma; \lambda; t)$ and their Sonnine partners at short time $t$ can be found by applying the following Tauberian theorem \cite{Feller}.
\begin{theorem}\label{th1}
{\rm
For $\hat{f}(s) = \mathscr{L}[f(t); s]$ such that for $s\to 0$ it can be presented as
\begin{equation*}%\label{30/08-4}
\hat{f}(s)  \simeq s^{-\rho} L\Big(\frac{1}{s}\Big), \quad \text{for} \quad \rho > 0, 
\end{equation*}
then 
\begin{equation*}%\label{30/08-5}
f(t) \simeq \frac{1}{\Gamma(\rho)} t^{\rho - 1} L(t), \qquad t\to\infty
\end{equation*}
for $L(t)$ being a slowly varying function at infinity, {\it i.e.}, $\lim_{t\to\infty}\frac{L(a t)}{L(t)} = 1$ for any $a > 0$.}
\end{theorem}
{\em Proof.}  The proof of this theorem is presented in Ref. \cite{Feller}. \qed

\smallskip
\noindent
Tauberian theorem is also valid if $s$ and $t$ are interchanges, that is $s\to\infty$ and $t\to 0$ \cite{Feller}.  
\begin{example}
{\rm The asymptotics of $\hat{k}_{1}(s)$ and $\hat{k}_{2}(\alpha, \gamma; \lambda; s)$ at $s\to 0$ read 
\begin{equation*}
\hat{k}_{1}(s) \simeq \frac{1}{s\ln(1/s)} \qquad \text{and} \qquad \hat{k}_{2}(\alpha, \gamma; \lambda; s) \simeq \lambda^{\gamma} \frac{s^{-(1 + \alpha\gamma)}}{\ln(1/s)}, 
\end{equation*}
for $\alpha, \gamma\in(0, 1)$ and $\lambda > 0$, which from Theorem \ref{th1} give
\begin{equation*}%\label{30/08-6}
k_{1}(t) \simeq \frac{1}{\ln t} \qquad \text{and} \qquad k_{2}(\alpha, \gamma; \lambda; t) \simeq \frac{t^{\,\alpha\gamma}}{\Gamma(1+\alpha\gamma)} \frac{1}{\ln t}, \qquad \text{for} \quad t\to \infty.
\end{equation*}
We note that at $s\to\infty$ we cannot apply Theorem \ref{th1} since $\hat{k}_{1}(s)$ and $\hat{k}_{2}(\alpha, \gamma; s)$ can be only approximated by $(\ln s)^{-1}$ without any power function.} 
\end{example}
\begin{example}
{\rm
The asymptotics of $\hat{M}_{1}(s)$ and $\hat{M}_{2}(\alpha, \gamma; \lambda; s)$ for small and large $s$ read, respectively
\begin{equation*}
\hat{M}_{1}(s) \simeq \ln(1/s) \quad \text{and} \quad \hat{M}_{2}(\alpha, \gamma; \lambda; s) \simeq s^{\alpha\gamma} \ln(1/s) \quad \text{for} \qquad s\to 0
\end{equation*}
as well as
\begin{equation*}
\hat{M}_{1}(s) = \hat{M}_{2}(\alpha, \gamma; \lambda; s) \simeq s^{-1} \ln s \qquad \text{for} \qquad s\to \infty. 
\end{equation*}
Then, we cannot apply Tauberian's theorem (Theorem \ref{th1}) because $\rho = 0$ and/or, respectively, $\rho = -\alpha\gamma$ which for $\alpha, \gamma\in(0, 1)$ gives negative value. For $s \gg 1$ the Theorem \ref{th1} gives 
\begin{equation*}%\label{1/09-8}
M_{1}(t) = M_{2}(\alpha, \gamma; \lambda; t) \simeq \ln\frac{1}{t}, \qquad \text{for} \quad t\to 0.
\end{equation*}
}
\end{example}

The use of the Laplace transform of convolution allows one to invert Eqs. \eqref{29/08-15} and \eqref{1/09-1}, as well as express the distributed order Prabhakar kernel and its Sonnine partner in the forms which correspond to $\hat{k}_1(s)$ and $\hat{M}_1(s)$. Indeed, we have
\begin{equation}\label{30/08-3}
k_{2}(\alpha, \gamma; \lambda; t) = \int_{0}^{t} k_{1}(t-\xi)\, e_{\alpha, 0}^{\,-\gamma}(\lambda; \xi) \D\xi
\end{equation}
and
\begin{equation}\label{30/08-3a}
M_{2}(\alpha, \gamma; \lambda; t) = \int_{0}^{t} M_{1}(t-\xi)\, e_{\alpha, 0}^{\,\gamma}(\lambda; \xi) \D\xi. 
\end{equation}
Moreover, the inverse Laplace transform of $\hat{k}_{1}(s)$ gives
\begin{equation}\label{2/09-1}
k_{1}(t) = \mathscr{L}^{-1}\left[\left(1 - \frac{1}{s}\right)\frac{1}{\ln s}; t\right] = \nu(t, -1) - \nu(t),
\end{equation}
where $\nu(t, p)$ and $\nu(t)$, for $p\in\mathbb{Z}$ and $t>0$, belong to the family of Volterra's function presented in Sec. \ref{sec2}. For $M_{1}(t)=\mathscr{L}^{-1}[\hat{M}_1(s); t]$ we take the inverse Laplace transform and use \cite[Eq. (2.5.2.2)]{APPrudikov-v5}. That allows one to get
\begin{equation}\label{1/09-10}
M_{1}(t) = \E^{t}\mathscr{L}^{-1}\left[\frac{\ln(s+1)}{s}; t\right] = - \E^{t} {\rm Ei}(-t),
\end{equation}
where ${\rm Ei}(-t) = -\int_{t}^{\infty}\exp(-u)/u\, \D u$ is the exponential integral. Next, we substitute Eqs. \eqref{2/09-1} and \eqref{1/09-10} into Eqs. \eqref{30/08-3} and \eqref{30/08-3a}, respectively. In the case of $k_{2}(\alpha, \gamma; t)$ Proposition \ref{prop2} enable us to write
\begin{equation*}%\label{30/08-11}
k_{2}(\alpha, \gamma; \lambda; t)  =  \sum_{n=0}^{\infty} \frac{(-\lambda)^{n}}{n!}(-\gamma)_{n} \left[\nu(t, \alpha n - 1) - \nu(t, \alpha n)\right],
\end{equation*}
where $\nu(t, \cdot)$ is Volterra's function and $(\gamma)_{n}$ is a Pochhammer (raising) symbol. The same result we obtain by taking the inverse Laplace transform of $\hat{k}_{2}(\alpha, \gamma; s)$ in which we first present $(1 + \lambda/s^{\alpha})^{\gamma}$ as the series indexed by $n\in\mathbb{N}_{0}$. The crucial step consists on calculating $\mathscr{L}^{-1}\{s^{-\alpha n - j} [\ln(s)]^{-1}; t\}$ with $j =0, 1$. That can be done by employing Eq. \eqref{eq:31082022-w11}. Moreover, Proposition \ref{prop2}b and Eq. \eqref{eq:31082022-w12} allow us to express $k_{2}(\alpha, \gamma; t)$ as the difference of Volterra-Prabhakar functions, %in the form %in terms of as the Fox $H$ function, namely
\begin{equation*}%\label{31/08-13}
k_{2}(\alpha, \gamma; \lambda; t) = \epsilon^{-\gamma}_{\alpha, -2}(\lambda; t) - \epsilon^{-\gamma}_{\alpha, -1}(\lambda; t).
\end{equation*}

To get the exact form of $M_{2}(\alpha, \gamma; t)$ we will use Eq. \eqref{30/08-3} for which we substitute Eq. \eqref{1/09-10} and apply the series form of the three parameters Mittag-Leffler function given by Eq. \eqref{eq:1092022-w11}. That gives
\begin{align}\label{3/09-10}
%\begin{split}
M_{2}(\alpha, \gamma; \lambda; t) & = - \int_{0}^{t} \E^{\xi} {\rm Ei}(-\xi)\, \sum_{r=0}^{\infty} \frac{(\gamma)_{r}\, (-\lambda)^{r}}{r!\, \Gamma(\alpha r)} (t - \xi)^{\alpha r - 1} \D\xi \nonumber\\
& = - \sum_{r=0}^{\infty} \frac{(\gamma)_{r}\, (-\lambda)^{r}}{r!\, \Gamma(\alpha r)}\, \int_{0}^{t} \E^{\xi} {\rm Ei}(-\xi)\, (t - \xi)^{\alpha r - 1} \D\xi,
%\end{split}
\end{align}
where we change (in a legitimate way) the order of sum and integral. The integral in Eq. \eqref{3/09-10} can be calculated by using \cite[Eq. (2.5.1.7)]{APPrudikov-v2} which implies
\begin{equation}\label{4/09-1}
\int_{0}^{t} \E^{\xi} {\rm Ei}(-\xi)\, (t - \xi)^{\alpha r - 1} \D\xi =  \Gamma(\alpha r)\, \ln(t)\, e_{1, 1+\alpha r}(t) - t^{\alpha r - 1} \sum_{j=1}^{\infty} t^{j}\frac{\psi(j+\alpha r)}{(\alpha r)_j},
\end{equation}
where $\psi(z) = \frac{\D}{\D z}\ln\Gamma(z)$ is the digamma function such that $\psi(1) = -C \simeq - 0.57721$ is equal to the Euler-Mascheroni constant taken with minus sign. Inserting it into Eq. \eqref{3/09-10} allows us to find the exact form of $M_{2}(\alpha, \gamma; t)$, namely
\begin{align}\label{4/09-2}
M_{2}(\alpha, \gamma; \lambda; t) & =  \ln\Big(\frac{1}{t}\Big) \sum_{r=0}^{\infty} \frac{(\gamma)_{r}}{r!} (-\lambda)^{r} e_{1, 1+\alpha r}(t) 
 + \sum_{r=0}^{\infty} \frac{(\gamma)_{r} (-\lambda t^{\alpha})^{r}}{r!\, \Gamma(\alpha r)}\, \sum_{j=1}^{\infty} t^{j-1}\frac{\psi(j+\alpha r)}{(\alpha r)_{j}}
\end{align}
in which we employ the series form of three parameters Mittag-Leffler function and Prabhakar function. At short time $t \ll 1$ from the asymptotic of three parameter Mittag-Leffler function we have $e_{1, 1+\alpha r}(t) \simeq t^{\alpha r}/\Gamma(1 + \alpha r)$ \cite{RGorenflo20}. That implies that the first term of Eq. \eqref{4/09-2} which contains the logarithmic function is proportional to $\ln(1/t) E_{\alpha, 1}^{\gamma}(-\lambda t^{\alpha})$ which for $t \ll 1$ gives $\ln(1/t)$. The second term with two series at short time vanishes because for $t\to 0$ the only term which can survive is for $r=0$ and $j=1$ that gives $\psi(1)/\Gamma(0) \sim 0$.
As a conclusion we can say that Eq. \eqref{4/09-2} reconstructs the behavior of $M_2(\alpha, \gamma; \lambda; t)$ found with the help of Tauberian theorem (Theorem \ref{th1}) at $t \ll 1$.

\section{Mean square displacement and higher moments}\label{sec4}

The PDF $p_{\hat{k}}(x, t) = \mathscr{L}^{-1}[\hat{p}_{\hat{k}}(x, s); t]$, where $\hat{p}_{\hat{k}}(x, s)$ is given by Eq. \eqref{29/08-12}, is symmetric function with respect to $x$. Thus, all its odd moments vanish and only the even ones are different from zero which for $n\in\mathbb{N}_{0}$ read
\begin{align}
    \begin{split}
    \langle x^{2n}(t)\rangle_{\hat{k}} & = \int_{\mathbb{R}} x^{2n} p_{\hat{k}}(x, t) \D x = \mathscr{L}^{-1}\left[\int_{\mathbb{R}} x^{2n} \hat{p}_{\hat{k}}(x, s) \D x; t \right] \\
    & = \mathscr{L}^{-1}\left[(-1)^n \frac{\D^{\,2n}}{\D\kappa^{2n}} \int_{\mathbb{R}} \E^{-\I\!\kappa x} \hat{p}_{\hat{k}}(x, s) \D x; t\right]_{\kappa = 0},
        \label{eq:30102022-w1}
    \end{split}
\end{align}
where we used $x^{2n} \exp(-\I\!\kappa x) = (-1)^n \frac{\D^{\,2n}}{\D\kappa^{2n}} \exp(-\I\!\kappa x)$. The integral in Eq. \eqref{eq:30102022-w1} is the Fourier transform of $\hat{p}_{\hat{k}}(x, s)$, so we can write
\begin{equation*}
\langle x^{2n}(t)\rangle_{\hat{k}} = \mathscr{L}^{-1}\left[(-1)^n \frac{\D^{\,2n}}{\D\kappa^{2n}} \tilde{\hat{p}}_{\hat{k}}(\kappa, s); t\right]
 %   \label{eq30102022-w2}
\end{equation*}
with $\tilde{\hat{p}}_{\hat{k}}(\kappa, s)$ given by Eq. \eqref{eq:20221006-w1a}. Setting now $y = \I\! \kappa\sqrt{B/[s \hat{k}(s)]}$ we get 
\begin{align*}
    \frac{\D^{\,2n}}{\D\kappa^{2n}}\,\tilde{\hat{p}}_{\hat{k}}(\kappa, s) & = (-1)^n \left[\frac{B}{s \hat{k}(s)}\right]^{n} \frac{1}{s} \frac{\D^{\,2n}}{\D y^{2n}} \frac{1}{1- y^2} \\
    & = (-1)^n \left[\frac{B}{s \hat{k}(s)}\right]^{n}\frac{1}{2 s}\frac{\D^{\,2n}}{\D y^{2n}}\left\{\frac{1}{1-y} +\frac{1}{1+y}\right\} 
    \\
    & = (-1)^n \left[\frac{B}{s \hat{k}(s)}\right]^{n}\frac{(2n)!}{2 s}\left\{\frac{1}{(1-y)^{2n+1}} +\frac{1}{(1+y)^{2n+1}}\right\}.
\end{align*}
Finally, at the limit of $y=0$, we obtain an extreme equal to 1.
\begin{align}
\begin{split}
\langle x^{2n}(t)\rangle_{\hat{k}} & = (2n)!\, B^n \mathscr{L}^{-1}\{s^{-1}[s\hat{k}(s)]^{-n}; t\}
= (2n)!\, B^n \mathscr{L}^{-1}[s^{-1}\hat{M}^n(s); t] \\
& = (2n)! B^n \int_{0}^{t} \mathscr{L}^{-1}[\hat{M}^n(s); u] \D u,
    \label{eq31102022-w1}
    \end{split}
\end{align}
where we apply the Sonnine equation \eqref{3/09-1}. 

All the moments $\langle x^{2n}(t)\rangle_{\hat{k}}$ for $\hat{k}(s) \equiv \hat{k}_1(s)$ and $\hat{k}(s) \equiv \hat{k}_2(\alpha, \gamma; \lambda; s)$ are non-negative which follows from the property (a1) given in \ref{app1}. Indeed, at the beginning of Sec. \ref{sec3} it is shown that $s \hat{k}_1(s)$ and $s \hat{k}_2(\alpha, \gamma; \lambda; s)$ are CBF. Thus, fdue to the property (a3) it appears that $[s\hat{k}_1(s)]^{-n}$ as well as $[s\hat{k}_2(\alpha, \gamma; \lambda; s)]^{-n}$ are CMFs. Next, the use of the property (a1) finished the proof.

\subsection{Mean square displacement}

Taking $n=1$ in Eq. \eqref{eq31102022-w1} we get the MSD 
\begin{equation*}%\label{4/09-5}
\langle x^{2}(t) \rangle_{\hat{k}} = 2 B \mathscr{L}^{-1}\{[s^{2} \hat{k}(s)]^{-1}; t\} = 2 B \int_{0}^{t} M(u) \D u.
\end{equation*}
The first approach to calculate the MSD for $\hat{k}_{1}(s)$ and $\hat{k}_{2}(\alpha, \gamma; s)$ is to use Tauberian theorem, see Theorem \ref{th1}. 
\begin{example}%\label{ex31102022-ex1}
{\rm For $\langle x^2(t)\rangle_{\hat{k}_1} \equiv \langle x^2(t)\rangle_{1}$ we have
\begin{equation*}%\label{4/09-6}
\mathscr{L}[\langle x^{2}(t)\rangle_{1}; s] =  2B [s^{2} \hat{k}_{1}(s)]^{-1} \simeq 2 B\left\{ {s^{-1} \ln(1/s), \quad \text{for} \quad s \to 0, \atop s^{-2} \ln s, \quad \text{for} \quad s\to\infty,} \right.
\end{equation*}
and respectively
\begin{equation*}%\label{4/09-7}
\langle x^{2}(t)\rangle_{1} \simeq 2 B\left\{ {\ln t, \quad \text{for} \quad t \to \infty, \atop t \ln(1/t), \quad \text{for} \quad t\to 0.} \right.
\end{equation*}}
\end{example}
\begin{example}%\label{ex310102022-ex2}
{\rm In the case of $\langle x^{2}(t) \rangle_{\hat{k}_2} \equiv \langle x^{2}(t) \rangle_{2}$ we can approximate it as follows
\begin{equation*}%\label{4/09-8}
\mathscr{L}[\langle x^{2}(t) \rangle_{2}; s] = 2B [s^{2} \hat{k}_{2}(\alpha, \gamma; s)]^{-1} 
\simeq 2B \left\{{\lambda^{-\gamma} s^{-(1-\alpha\gamma)} \ln(1/s), \quad \text{for} \quad s\to 0, \atop s^{-2} \ln s, \qquad \text{for} \quad s\to\infty.} \right.
\end{equation*}
Because $\alpha\gamma\in(0, 1)$ then $1 - \alpha\gamma\in(0, 1)$ and we can apply the Tauberian theory which gives
\begin{equation}\label{4/09-9}
\langle x^{2}(t) \rangle_{2} \simeq 2B \left\{{\lambda^{-\gamma} \frac{t^{-\alpha\gamma}}{\Gamma(1-\alpha\gamma)} \ln t, \quad \text{for} \quad t\to\infty, \atop t \ln(1/t), \qquad \text{for} \quad t\to 0.} \right.
\end{equation}}
\end{example}

In the next approach we calculate the exact form of $\langle x^{2}(t) \rangle_{1}$ and $\langle x^{2}(t) \rangle_{2}$. For $\langle x^{2}(t) \rangle_{1}$ we have
\begin{align}\label{4/09-11}
\begin{split}
\langle x^{2}(t) \rangle_{1} & = 2B \int_{0}^{t} M_{1}(u) \D u =  - 2B\int_{0}^{t} \E^{u} {\rm Ei}(-u) \D u \\
& = 2B \sum_{j=1}^{\infty} \frac{t^{j}}{j!}[\psi(1 + j) - \ln t] \\
& = 2B\sum_{j=1}^{\infty} \frac{t^{j}}{j!} \psi(1+j) + 2B (\E^{t} - 1)\ln\Big(\frac{1}{t}\Big),
\end{split}
\end{align}
where we employ \cite[Eq. (2.5.1.7)]{APPrudikov-v2}. The series in Eq. \eqref{4/09-11} is the exponential generating function which is given by \cite[Eq. (4.2a)]{ARMiller06} and reads
\begin{equation}\label{6/09-5}
\sum_{j=1}^{\infty} \frac{t^{j}}{j!} \psi(1+j) = C + \E^{t}[\ln(t) - {\rm Ei}(-t)].
\end{equation}
Eq. \eqref{6/09-5} enables us to express $\langle x^{2}(t) \rangle_{1}$ in the form
\begin{equation}\label{eq6092022-w6}
\langle x^{2}(t) \rangle_{1} = 2B[C + \ln(t) - \E^{t} {\rm Ei}(-t)].
\end{equation}
Let us now consider its asymptotic behavior at short and long time $t$. At $t \ll 1$ we can approximate $E_{1}(t) = -{\rm Ei}(-t) \simeq -C - \ln(t) + t$ \cite[Eq. (6.6.2)]{NIST}. That reduces Eq. \eqref{eq6092022-w6} into  $\langle x^{2}(t) \rangle_{1} \simeq 2B \left[\left(-C + \ln\frac{1}{t}\right) (\E^{t}-1) + t\E^{t}\right]$.
Next, notice that $\exp(t) \simeq 1 + t$ at short $t$ we approximate it as
\begin{equation*}
    \langle x^{2}(t) \rangle_{1} \simeq 2B (1- C)t + 2B t\ln\frac{1}{t} \simeq 2B t\ln\frac{1}{t},
\end{equation*}
this is the asymptotics obtained from Theorem \ref{th1}. At $t \gg 1$ the exponential integral behaves as $E_{1}(t) = -{\rm Ei}(-t) \simeq \exp(-t)/t$ \cite[Eq. (6.12.1)]{NIST}. It means that for $t\gg 1$ we have $\langle x^{2}(t) \rangle_{1} \simeq 2B [C + \ln(t) - t^{-1}]$. The most important term there is equal to $2B\ln(t)$ which was indicated by Tauberian theorem (Theorem \ref{th1}).

In the case of the MSD calculated for the distributed order Prabhakar derivative we proceed similarly as for $\langle x^2(t)\rangle_1$. Hence, we can write
\begin{align*}%\label{6/09-7}
%\begin{split}
\langle x^{2}(t) \rangle_{2} & = 2 B \int_{0}^{t} M_{2}(\alpha, \gamma; \lambda; u) \D u = 2B \int_{0}^{t} \left\{\int_{0}^{u}M_{1}(\xi)\,  e_{\alpha, 0}^{\gamma}(\lambda; u -\xi) \D\xi\right\} \D u \\
& = - \sum_{r=0}^{\infty} \frac{(\gamma)_{r} (-\lambda)^{r}}{r! \Gamma(\alpha r)} \int_{0}^{t} \left[\int_{0}^{u} \E^{\xi} {\rm Ei}(-\xi) (u - \xi)^{\alpha r - 1} \D\xi \right]\D u,
%\end{split}
\end{align*}
where in the first equality we apply Eq. \eqref{30/08-3a}. To get the next equality we employed the series form of three parameter Mittag-Leffler function \eqref{eq:1092022-w11}. The integral over $\xi\in(0, u)$ inside the square braced is calculated in Eq. \eqref{4/09-1}. That allows one to present $\langle x^{2}(t) \rangle_{2}$ as
\begin{equation}\label{7/09-1}
\langle x^{2}(t) \rangle_{2}  = -2 B \sum_{r=0}^{\infty} \frac{(\gamma)_{r} (-\lambda)^{r}}{r!} \left[\int_{0}^{t} \ln(u)\, e_{1, 1+\alpha r}(1; u) \D u + \sum_{j\geq 1} \frac{\psi(j + \alpha r)}{\Gamma(j + \alpha t)}\int_{0}^{t} u^{\alpha r + j - 1} \D u\right]\!.
\end{equation}
For calculation of the first integral in the square bracket we need the following Proposition \ref{7/09-1p}.
\begin{proposition}\label{7/09-1p}
{\rm 
For $a > 0$ and $t > 0$ we have
\begin{equation}\label{7/09-5}
\int_{0}^{t} \ln(u)\, e_{1, 1 + a}(1; u)\D u = \ln(t)\, e_{1, 2 + a}(t) - \frac{t^{1 + a}}{1 + a} \frac{1}{\Gamma(2+a)} {_{2}F_{2}}\left({1, a + 1 \atop a + 2, a + 2}; t\right).
\end{equation}
}
\end{proposition}
{\em Proof. } Let us employ the series form of the three parameters Mittag-Leffler function \eqref{eq:1092022-w11}. Hence, we have
\begin{align}\label{7/09-6}
    \begin{split}
        \int_{0}^{t} \ln(u)\, e_{1, 1+a}(u) \D u & = \sum_{j=0}^{\infty} \frac{1}{\Gamma(1 + j + a)} \int_0^t u^{a+j} \ln(u) \D u \\
        & = \ln(t)\, \sum_{j=0}^{\infty} \frac{t^{1 + j + a}}{\Gamma(2 + j + a)} - t^{1 + a} \sum_{j\geq 0} \frac{t^j}{(1 + j + a)\Gamma(2 + j + a)},
    \end{split}
\end{align}
where we used \cite[Eq. (2.6.2.2.)]{APPrudnikov-v1} according to which 
\begin{equation*}
\int_{0}^{t} u^{a + 1} \ln(u) \D u = \frac{t^{1 + j + a}}{1+j+a}\Big[\ln(t) - (1 + j + a)^{-1}\Big].
\end{equation*}
Thereafter, the first series in Eq. \eqref{7/09-6} due to Eq. \eqref{eq:1092022-w11} is the Prabhakar function. The second series there gives the generalized hypergeometric function of type ${_{2}F_{2}}$. That finishes the proof. \qed

\medskip
\noindent
The second integral in the square bracket reads 
\begin{equation}\label{8/09-1}
    \int_0^t u^{\alpha r + j - 1} \D u = \frac{t^{\alpha r + j}}{\alpha r + j}.
\end{equation}
Finally, inserting Eqs. \eqref{7/09-5} and \eqref{8/09-1} into Eq. \eqref{7/09-1} we obtain
\begin{align}\label{8/09-2}
\begin{split}
    \langle x^2(t)\rangle_{2} & = \ln\left(\frac{1}{t}\right) \sum_{r=0}^{\infty} \frac{(\gamma)_{r}\, (-\lambda)^r}{r!}\, e_{1, 2 + \alpha r}(1; t) \\
    & + t \sum_{r=0}^{\infty} \frac{(\gamma)_{r}\, (-\lambda t^{\alpha})^r}{(1 + \alpha r) r!\, \Gamma(2 + \alpha r)}\, {{_2}F_{2}}\left({1, 1 + \alpha r\atop 2 + \alpha r, 2 + \alpha r}; t\right) \\
    & + \sum_{r=0}^{\infty} \frac{(\gamma)_r (-\lambda t^\alpha)^r}{r!\, \Gamma(\alpha r)} \sum_{j= 1}^{\infty}  t^j\,\frac{\psi(j + \alpha r)}{(\alpha r)_j}.
\end{split}    
\end{align}
In a short time, the second and third series in Eq. \eqref{8/09-2} can be neglected. At $t \ll 1$ we can approximate $e_{1, 2+\alpha r}(1; t)$ as $t^{1+\alpha r}/\Gamma(2 + \alpha r)$ and insert it into the first series. That gives $t\ln(1/t) E_{\alpha, 2}^{\gamma}(-\lambda t^{\alpha})$ where $E_{\alpha, 2}^{\gamma}(-\lambda t^{\alpha})\sim 1/\Gamma(2) = 1$. Thus, $\langle x^2(t)\rangle_{2}$ at $t \ll 1$ is proportional to $t\ln(1/t)$. It means that we obtain the behavior indicated by Tauberian theorem (Theorem \ref{th1}), see Eq. \eqref{4/09-9}. In the opposite case, {\it i.e.}, at large $t$, the most important term is also the first series. From \cite[Eq. (4.4.17)]{RGorenflo20} appears that $e_{1, 2+\alpha r}(1; t) \simeq -t^{\alpha r}/\Gamma(1 + \alpha r)$ for $t\gg 1$. Hence, $\langle x^2(t)\rangle_{2} \simeq \ln(t) E_{\alpha, 1}^\gamma(-\lambda t)$,
in which with the help of \cite{RGarra18} we approximate $E_{\alpha, 1}^\gamma(-\lambda t)$ as $\lambda^{-\gamma} t^{-\alpha\gamma}/\Gamma(1-\alpha\gamma)$. That leads us to the asymptotic behavior of $\langle x^2(t)\rangle_{2}$ at the large time  given by Tauberian theorem, see Eq. \eqref{4/09-9}.

\subsection{Standard deviation, skewness, and kurtosis}

Because the odd moments are equal to zero then the standard deviation reads
\begin{equation*}%\label{eq2112022-w1}
\sigma = \sqrt{\langle x^2(t)\rangle_{\hat{k}}}.
\end{equation*}
The MSD $\langle x^2(t)\rangle_{\hat{k}_1} \equiv \langle x^2(t)\rangle_{1}$ is given by Eq. \eqref{eq6092022-w6} whereas $\langle x^2(t)\rangle_{\hat{k}_2} \equiv \langle x^2(t)\rangle_{2}$ is presented by Eq. \eqref{8/09-2}. Moreover, the skewness $\mu_3 = \mathbb{E}[(x - \langle x\rangle)^3]/\sigma^3$ proportional to the odd moments vanishes which appears for symmetric PDF. The expectation operator is denoted by $\mathbb{E}$.

The kurtosis according to definition for symmetric PDF figures out
\begin{equation*}%\label{eq2112022-w2}
\mu_4 = \frac{\mathbb{E}[(x - \langle x \rangle_{\hat{k}})^4]}{\sigma^4} = \frac{\langle x^4 \rangle_{\hat{k}}}{\langle x^2 \rangle_{\hat{k}}^2}.
\end{equation*}
Eq. \eqref{eq31102022-w1} for $n = 2$ gives the 4th moments which depends on the MSD. Indeed, we have
\begin{align}\label{eq2112022-w3}
\begin{split}
    \langle x^4(t) \rangle_{\hat{k}} & = 24 B^2 \int_0^t \mathscr{L}^{-1}[\hat{M}^2(s); u] \D u \\
    & = 24 B^2 \int_0^t \left[\int_0^u M(u-\xi) M(\xi) \D\xi\right]\D u \\
    & = 24 B^2 \int_0^t M(\xi) \left[\int_{\xi}^t M(u - \xi) \D u \right]\D\xi.
    \end{split}
\end{align}
Setting now $u - \xi = y$ we get that the integral in square bracket reads
\begin{equation*}%\label{eq2112022-w4}
        \int_{\xi}^t M(u - \xi) \D u = \int_0^{t-\xi} M(y) \D y = (2 B)^{-1} \langle x^2(t-\xi)\rangle_{\hat{k}}.
\end{equation*}
Substituting this results in Eq. \eqref{eq2112022-w3} we obtain
\begin{equation}\label{eq2112022-w5}
\langle x^4(t) \rangle_{\hat{k}} = 12 B \int_0^t M(\xi)\, \langle x^2(t-\xi)\rangle_{\hat{k}}\D\xi.
\end{equation}

\begin{proposition}\label{28112022-p1} 
{\rm
For $t>0$ the following convolution integral formula holds true
\begin{align*}
\int_0^t \E^{\xi} {\rm Ei}(-\xi) \ln(t-\xi) \D\xi & = - C^2 \E^t + \frac{\pi^2}{6} ( 1 - \E^t) - (2 C + \ln t) \E^t (\ln t) + (C + 2\ln t) \E^t {\rm Ei}(-t) \\
& - C (\ln t) - (\ln t)^2 + 2 t \E^t {_3 F_3}\left({1, 1, 1 \atop 2, 2, 2}; -t\right).
\end{align*}
}
\end{proposition}
\noindent
{\em Proof.} By Laplace transform method and convolution theorem, we have
\begin{align}\label{23Dec2022-1}
\begin{split}
    \mathscr{L}\left[\int_0^t \E^{\xi} {\rm Ei}(-\xi) \ln(t-\xi) \D\xi; s\right] & = \mathscr{L}\left[\E^{t}{\rm Ei}(-t); s\right]\,\mathscr{L}\left[\ln(t); s\right] \\
    & = - \mathscr{L}[{\rm Ei}(-t); s-1]\, \frac{1}{s}(C + \ln s)\\
    & =\frac{\ln^2 s}{s(s-1)}+C\frac{\ln s}{s(s-1)},
    \end{split}
\end{align}
where we have used Eq. \eqref{1/09-10} and $\mathscr{L}[\ln t; s] = -(C + \ln s)/s$. The first inverse Laplace transform can be calculated by using \cite[Eq. (1.1.1.13)]{APPrudikov-v5}, namely 
\begin{equation*}
\mathscr{L}^{-1}\left[\frac{F(s)}{s+a}; t\right] = \E^{-a t} \int_0^t \mathscr{L}^{-1}[F(s); u] \E^{a u} \D u.
\end{equation*}
Setting $a = -1$ and $F(s) = (\ln s)^2/s$, we have
\begin{align}\label{23/12-1}
\begin{split}
\mathscr{L}^{-1}\left[\frac{\ln^2 s}{s(s-1)}; t\right] & = \E^{t} \int_0^t \E^u \mathscr{L}^{-1}\left[\frac{(\ln s)^2}{s}; u\right] \D u \\
%& = \E^{t} \int_0^t \E^u \left[(C + \ln u)^2 - \frac{\pi^2}{6}\right] \D u\\
& = \left(C^2 - \frac{\pi^2}{6}\right) \E^t \int_0^t \E^{-u} \D u + 2C \E^t \int_0^t \E^{-u}(\ln u) \D u + \E^t \int_0^t \E^{-u} (\ln u)^2 \D u \\
& = \left(C^2 - \frac{\pi^2}{6}\right)(\E^t-1) + 2C \E^t \int_0^t \E^{-u}(\ln u) \D u + \E^t \int_0^t \E^{-u} (\ln u)^2 \D u,
\end{split}
\end{align}
where according to \cite[Eq. (2.5.1.7)]{APPrudikov-v5}, we have
\begin{equation}\label{23Dec2022-3}
    \mathscr{L}^{-1}\left[\frac{\ln^2 s}{s}; t \right] = (C + \ln t)^2 - \frac{\pi^2}{6}.
\end{equation}
To calculate the first integral given by Eq. \eqref{23/12-1} we can apply \cite[Eq. (1.6.10.2)]{APPrudnikov-v1} which leads us to
\begin{equation*}
\int_0^t \E^{-u} (\ln u) \D u = {\rm Ei}(-t) - C - \E^{-t}\ln t.
\end{equation*}
In the next integral we use the series form of exponential function and change (in legitimate way) the order of integration and series. That allows us to get
\begin{align*}
    \begin{split}
        \int_0^t \E^{-u} (\ln u)^2 \D u & = \sum_{r=0}^{\infty} \frac{(-1)^r}{r!} \int_0^t u^r (\ln u)^2 \\
        & = 2 t \sum_{r=0}^{\infty} \frac{(-t)^r}{r! (1+r)^3} - 2 t (\ln t) \sum_{r=0}^{\infty} \frac{(-t)^r}{r! (1+r)^2} - (\ln t)^2 \sum_{r=0}^{\infty}\frac{(-t)^{1+r}}{(1+r)!} \\
        & = 2 t\, {_3 F_3}\left({1, 1, 1 \atop 2, 2, 2}; -t\right) - 2 t\, (\ln t)\, {_2 F_2}\left({1, 1\atop 2, 2}; -t\right) - (\ln t)^2 (\E^{-t} - 1) \\
        & = 2 t\, {_3 F_3}\left({1, 1, 1 \atop 2, 2, 2}; -t\right) - 2 (\ln t) (C -{\rm Ei}(-t) + \ln t) - (\ln t)^2 (\E^{-t} - 1),
    \end{split}
\end{align*}
where ${_2 F_2}\big({1, 1 \atop 2, 2}; -t\big) = (C -{\rm Ei}(-t) + \ln t)/t$. From \cite[Eq. (1.25)]{HMSrivastava13} we can obtain that ${_3 F_3}\big({1, 1, 1 \atop 2, 2, 2}; -t\big)$ is equal to $\Phi^{\star (1, 1)}_{1; 1}(-t, 3, 1)$. Finally, we have
\begin{align}\label{23/12-10}
\begin{split}
\mathscr{L}^{-1}\left[\frac{\ln^2 s}{s(s-1)}; t\right] & = - \left(C^2 + \frac{\pi^2}{6}\right) \E^t + 2 t \E^t {_3 F_3}\left({1, 1, 1 \atop 2, 2, 2}; -t\right) + 2(C +\ln t) \E^t {\rm Ei}(-t) \\& - (2 C + \ln t) (\E^t + 1)(\ln t) - C^2 + \frac{\pi^2}{6}.
\end{split}
\end{align}
The latter inverse Laplace transform reads
\begin{align}\label{23Dec2022-4}
    \mathscr{L}^{-1}\left[\frac{\ln s}{s(s-1)}; t\right] = \int_0^t \mathscr{L}^{-1}\left[\frac{\ln s}{s-1}; u\right] \D u  = \int_0^t M_1(u) \D u = \frac{\langle x^2(t) \rangle_1}{2B},
\end{align}
where we employed Eqs. \eqref{1/09-10} and \eqref{4/09-11}. Insertion of Eqs. \eqref{23/12-10} and \eqref{23Dec2022-4} into the inverse Laplace transform of Eq. \eqref{23Dec2022-1} finishes the proof. \qed

\begin{example}%\label{ex5112022-1}
{\rm From Eqs. \eqref{eq2112022-w5} and \eqref{eq6092022-w6}, we obtain %Eq. \eqref{eq2112022-w5} for $\hat{k}_{1}$ whose Sonnine partner is $\hat{M}_{1}(s)$, this is after inserting to it $\langle x^2(t-\xi)\rangle_{1}$ given by Eq. \eqref{eq6092022-w6}, we obtain 
\begin{align}
    \begin{split}
    \langle x^4(t)\rangle_{1} & = 24 B^2 \int_0^t M_1(\xi)\big[C  - \E^{t-\xi}{\rm Ei}(\xi-t) + \ln(t-\xi)\big] \D\xi \\
   & = 12 B C\, \langle x^2(t)\rangle_1 + 24 B^2 
    \E^{t} \int_0^t {\rm Ei}(-\xi) {\rm Ei}(\xi - t) \D\xi - 24 B^2 \int_0^t \E^{\xi} {\rm Ei}(-\xi) \ln(t-\xi) \D\xi.
        \label{eq5112022-w1}
    \end{split}
\end{align}

The second term in Eq. \eqref{eq5112022-w1} can be calculated by using \cite[Eq. (2.5.11.6) p. 77]{APPrudikov-v2} and it reads
\begin{align*}
    \int_0^t {\rm Ei}(-\xi) {\rm Ei}(\xi - t) \D\xi & = 2(C +\ln t)\E^{-t} - 2 (1 - t C - t \ln t) {\rm Ei}(-t) - t \left[\frac{\pi^{2}}{6} + (C +\ln t)^{2}\right] \\ 
    & +2 t^{2} {_{3}F_{3}}\left({1, 1, 1 \atop 2, 2, 2}; -t\right).
    \end{align*}
For the integral of the third term, we will apply the last proposition \ref{28112022-p1}. That allows us to obtain the exact form of $\langle x^4(t)\rangle_{1}$.
}
\end{example}

\section{Probability density function $p_{\hat{k}}(x, t)$}\label{sec5}

We begin with finding the explicit form of the PDF $p_{\hat{k}}(x, t)$ for the memory kernels $k_1(t)$ and $k_2(\alpha, \gamma; \lambda; t)$ for which we have $p_{\hat{k}_1}(x, t)\equiv p_1(x, t)$ and $p_{\hat{k}_2}(x, t)\equiv p_2(x, t)$, respectively. For that purpose we are using the general form of PDF given by Eq. \eqref{29/08-12} in which we take $\hat{k}_1(s)$ from Eq. \eqref{3/09-2} and $\hat{k}_2(\alpha, \gamma; \lambda; s)$ from Eq. \eqref{29/08-15}. Thus, we have %at allows us to write
\begin{equation}
    p_{1}(x, t) = \sum_{r=0}^{\infty} p_{1}(r; x, t) \quad \text{where} \quad p_{1}(r; x, t) = 
    \frac{1}{2\sqrt{B}} %\sum_{r=0}^{\infty} 
    \left(\frac{-|x|}{\sqrt{B}}\right)^r\, \frac{1}{r!\, \Gamma(b_r)}\, \epsilon_{1,b_r - 1, -b_r}^{-b_r}(t),
    \label{eq:20220901-w4}
\end{equation}
where $b_r = (r+1)/2$, as well as
\begin{equation}
p_{2}(x, t) = \sum_{r=0}^{\infty}\int_0^{\infty} p_{1}(r; x, t-\xi)\, e_{1, 0}^{-\gamma b_r}(\lambda, \xi) \D\xi.
%=\frac{1}{2\sqrt{B}} \sum_{r=0}^{\infty} \left(\frac{-|x|}{\sqrt{B}}\right)^r\, \frac{1}{r!\, \Gamma(b_r)}\, \int_0^{\infty} \epsilon_{1, b_r - 1, -b_r}^{-b_r}(t-\xi)\, e_{1, 0}^{-\gamma b_r}(\lambda, \xi) \D\xi,
    \label{eq:20221008-w5}
\end{equation}
%where $b_r = (r+1)/2$.
\\
\noindent
{\em Proof of Eqs. \eqref{eq:20220901-w4} and \eqref{eq:20221008-w5}. } Without specifying the memory function $\hat{k}(s)$ we express Eq. \eqref{29/08-12} as
\begin{equation}\label{22/09-3}
    %p_{\hat{k}}(x, t) = \sum_{r=0}^{\infty} p_{\hat{k}}(r; x, t) %& = \frac{1}{2\sqrt{B}} \mathscr{L}^{-1}\left\{\sum_{r\geq0} \frac{(-|x|/\sqrt{B})^r}{r!} s^{\frac{r-1}{2}}\, [\hat{k}(s)]^{\frac{r+1}{2}}; t\right\}\\
   % \quad \text{where} \quad
    p_{\hat{k}}(x, t) = \frac{1}{2\sqrt{B}} \sum_{r=0}^{\infty} \frac{(-|x|/\sqrt{B})^r}{r!}\, \mathscr{L}^{-1}\left\{s^{b_r - 1}\, [\hat{k}(s)]^{b_r}; t\right\}
\end{equation}
and we set $b_r = (r+1)/2$. Here, we assume that the inverse Laplace transform of the series is the same as the series of inverse Laplace transforms. Next, we express $[\hat{k}_1(s)]^{b_r}$ in the series form 
\begin{equation*}%\label{22/09-4}
    [\hat{k}(s)]^{b_r} = (\ln s)^{-b_r}\,\left(1 - \frac{1}{s}\right)^{b_r} = (\ln s)^{-b_r}\, \sum_{j=0}^{\infty} \frac{s^{-j}}{j!} (-b_r)_{j}.
\end{equation*}
That allows one to calculate $\mathscr{L}^{-1}[s^{b_r-1}\, \hat{k}^{b_r}_1(s); t]$, namely
\begin{align}
\begin{split}
\mathscr{L}^{-1}[s^{b_r-1}\, \hat{k}^{b_r}_1(s); t] %& = \mathscr{L}^{-1}\left[s^{b_r-1} \Big(1-\frac{1}{s}\Big)^{b_r} \ln^{-b_r}(s); t\right]\\
& = \sum_{j=0}^{\infty} \frac{(-b_r)_j}{j!} \mathscr{L}^{-1}\left[s^{b_r -j -1} (\ln s)^{-b_r}; t\right] \\
& = \sum_{j=0}^{\infty} \frac{(-b_r)_j}{j!}\, \mu(t, b_r-1, j-b_r)\\
& = \frac{1}{\Gamma(b_r)} \epsilon^{-b_r}_{1, b_r - 1, -b_r}(t),
    \label{eq:20221007-w10}
    \end{split}
\end{align}
where we used Proposition \ref{p:20221009-p1}. %the integral form of Volterra's function given by Eq. \eqref{eq:31082022-w10} 
%we get 
%\begin{equation}
%\mathscr{L}^{-1}[s^{b_r-1}\, \hat{k}^{b_r}_1(s); t] = \frac{1}{\Gamma(b_r)} \epsilon^{-b_r}_{1, b_r - 1, -b_r}(t).
%    \label{eq:20221007-w11}
%\end{equation}
Inserting it into Eq. \eqref{22/09-3} we end up the proof of Eq. \eqref{eq:20220901-w4}. 

In the case of $p_2(x, t)$ using the memory function $k_2(\alpha, \gamma; \lambda; s)$ only the inverse Laplace transform in Eq. \eqref{22/09-3} will be changed. Employing Eq. \eqref{29/08-15} and the Laplace transform of convolution we have
\begin{align*}
%\begin{split}
\mathscr{L}^{-1}\{s^{b_r - 1} [\hat{k}_2(\alpha, \gamma; \lambda; s)]^{b_r}; t\} & = \mathscr{L}^{-1}\left[s^{b_r - 1} \hat{k}_1^{b_r}(s)\, \Big(1 + \frac{\lambda}{s^\alpha}\Big)^{\gamma b_r}; t\right] \\
& = \int_0^{\infty} \mathscr{L}^{-1}[s^{b_r - 1} \hat{k}_1^{b_r}(s), t-\xi]\, e_{\alpha, 0}^{-\gamma b_r}(\lambda; \xi) \D\xi.
    \label{eq:20221008-w6}
%    \end{split}
\end{align*}
Employing now Eq. \eqref{eq:20221007-w10}, we finish the proof of Eq. \eqref{eq:20221008-w5}. \qed

%As is shown in Eq. \eqref{29/08-12} $\hat{p}(x, s)$ is symmetric with respect to $x$ from this reason all odd moments vanishes. Thus, the skewness $\mu_3$ {\color{red} proportional to the third and first moments} is equal to zero. For the even moments we have
%\begin{align}
%\langle x^{2m}(t) \rangle_{\hat{k}} & = \int_{-\infty}^{\infty} x^{2m}\, p_{\hat{k}}(x, t) \D x = \mathscr{L}^{-1}\left[\int_{-\infty}^{\infty} x^{2m}\, \hat{p}(x, s) \D s\right] \\
%& = \mathscr{L}^{-1}\left[\int_{-\infty}^{\infty} (-1)^m \frac{\D^{\,2m}}{\D \kappa^{2m}} \E^{\I\!\kappa x} \hat{p}_{\hat{k}}(x, s) \D x\right]_{\kappa = 0},
%    \label{eq:25102022-w1}
%\end{align}
%where we used the fact that $x^{2m} = (-1)^m \frac{\D^{\,2m}}{\D\kappa^{2m}} \E^{\I\!\kappa x}\lvert_{\kappa=0}$.

\section{Conclusion}\label{sec6}

The paper presents the exact solution of the generalized Fokker-Planck equation whose integral kernel $k(t)$ informs on the form of smearing first-time derivative and it is called a memory function. With respect to the form of the memory function, we considered two examples of the generalized Fokker-Planck equation. In one of them, we take the distributed order kernel denoted by $k_1(t)$. In the next example, we studied the generalized Fokker-Planck equation with distributed order Prabhakar kernel $k_2(\alpha, \gamma; \lambda; t)$. We showed that $k_1(t)$ and $k_2(\alpha, \gamma; \lambda; t)$ can be expressed as the differences of Volterra functions or Volterra-Prabhakar functions, which was introduced in the present paper, respectively. Thereafter, we calculated the MSDs, 4th moments, and PDFs. The non-negativity of these functions is ensured by the Stieltjes character of $k_1(t)$ and $k_2(\alpha, \gamma; \lambda; t)$.

\section*{Acknowledgments}

KG and TP research was supported by the NCN Research Grant Preludium Bis 2 No. UMO-2020/39/O/ST2/01563. TS acknowledges financial support by the German Science Foundation (DFG, Grant number ME 1535/12-1) and by the Alexander von Humboldt Foundation. Ž.T. was supported by the Department of Mathematics, Faculty of Sciences, University of Ostrava.

\appendix

\section{Completely monotone, Stieltjes, Bernstein, and completely Bernstein functions - a brief tutorial \cite{RLSchilling12, TS_ZT_book}}\label{app1}

The completely monotone functions (CMFs) are a class of non-negative functions $G(s)$ of a non-negative argument whose all derivatives exist for $s>0$ and  alternate, {\it i.e.}, $(-1)^{n}\, G^{(n)}(s) \geq 0$, $n \in\mathbb{N}_0$, where $G^{(n)}(s) = \D^{\,n} G(s)/ \D s^{n}$. 
According to the Bernstein theorem \cite{RLSchilling12} we can connect in a unique way the CMF and non-negative functions: $s\in [0,\infty)\rightarrow G(s) \in \textrm{CMF}$ iff  
\begin{equation}\label{17/06-1}
G(s) = \int_{0}^{\infty} \exp(-s t)\, g(t)\!\D t 
\end{equation} 
and if $g(t)\ge 0$ for all $t\in [0,\infty)$.  
\\
Among the important properties of CMFs we present the following.
\begin{itemize}
\item[(a1)] The product of two CMFs is also CMF.
\end{itemize}
Note that Eq. \eqref{17/06-1} is the real-valued Laplace integral of a non-negative function $g(t)$ and in order to deal with the Laplace transform its argument has to be complex. Hence, $G(s)$ and the Laplace transform of $g(t)$, denoted as $\hat{g}(z)$ are different objects: the first of them is a real function of $s > 0$ while the second is complex-valued and depends on $z \in \mathbb{C}\setminus\mathbb{R}_{-}$. Knowledge of analytic continuation of $G(s)\rightarrow\hat{g}(z)$ is important  because these are special analytic properties of $\hat{g}(z)$ (known as Herglotz conditions, \cite{Berg, Akhiezier}) which  determine, e.g., according to the Theorem 2.6 of Ref.~\cite{GGripenberg90} quoted as Theorem of Ref.~\cite{ECapelas11}, conditions under which $\hat{g}(z)$ is representable as the Laplace transform of a nonnegative measure defined on positive semi-axis. In the majority of probabilistic applications, we may restrict considerations to the real variable $s$ and treat $\hat{g}(s)$ as the real function $G(s)$ but to find the inverse Laplace transform of $\hat{g}(z)$ we must have the variable $z$ complex.

The next class of functions needed in our considerations is that of complete Bernstein functions (CBF) \cite{RLSchilling12, TS_ZT_book}: $c(s)$ is CBF, $s > 0$, if $c(s)/s$ is the Laplace transform of CMF restricted to the positive semiaxis, or, equivalently, in the same way restricted Stieltjes transform of a positive function named also as the Stieltjes function (SF). Note that all SFs are completely monotone, {\it i.e.}, SFs are a subclass of CMF. 

The following properties of CBFs and connected to them SFs will be used:
\begin{itemize}
\item[(a2)] The composition of CBFs is CBF; 
\item[(a3)] The composition of CMF and CBF is another CMF;
\item[(a4)] The composition of SF and CBF as well as CBF and SF is another SF;
\item[(a5)] The algebraical inversion of SF is CBF and the algebraical inversion of CBF is SF;
\item[(a6)] If $c(s)$ is CBF then $c(s)/s$ is SF and if $c(s)$ is SF then $s/c(s)$ is CBF.
\end{itemize}

\smallskip
\noindent
CBFs form a subclass of the Bernstein functions (BF). These are defined as non-negative functions whose derivative is CMF \cite{RLSchilling12, TS_ZT_book}: $h(s) > 0$ is BF if
\begin{equation*}
(-1)^{n-1}\, h^{(n)}(s) \geq 0, \quad n = 1, 2,\dots.
\end{equation*}

%The example which illustrates similarities and differences between CMF, SF, BF, and CBF is the power function. This is presented in Tab. \ref{tab1}.
%\begin{table}
%\begin{center}
%\begin{tabular}{c | c | c | c| c}
%$f(s)$\,\,\, & \,\,\,CMF\,\,\, & \,\,\,SF\,\,\, & \,\,\,BF\,\,\, & \,\,\,CBF\,\,\, \\ \hline 
%$s^{\mu}$ & $\mu \leq 0$ & $\mu \in(-1, 0)$ & $\mu\in[0, 1]$ & $\mu\in(0, 1)$ \\
%$(s + b)^{\mu}$, $b > 0$ & $\mu \leq 0$ & $\mu \in(-1, 0)$ & $\mu\in[0, 1]$ & $\mu\in(0, 1)$ 
%\end{tabular}
%\caption{\label{tab1} Examples of CMF, SF, BF, and CBF.}
%\en
d{center}
%\end{table}

\section{Proposition \ref{28112022-p1} for $t \ll 1$}

\noindent
For $t\in(0, 1)$ the following convolution integral formula holds true
\begin{align*}% \label{12/12-22}
%\begin{split}
\int_0^t \E^{\xi} {\rm Ei}(-\xi) \ln(t-\xi) \D\xi & = -\E^t (C + \ln t)^2 + C \E^t {\rm Ei}(-t) + 2(\ln t) \E^t {\rm Ei}(-t) - C (\ln t) - (\ln t)^2 \\
&  - \sum_{n=1}^{\infty}\frac{1}{(n+1)^2}\sum_{r=1}^{n}\frac{t^r}{r!} + \E^t\int_{0}^{t}\left\{\frac{2}{u}\,{\rm Ein}(u) + \frac{\E^{-u}}{u}\,{\rm Ein}(-u)\right\} \D u,  
\end{align*} 
where ${\rm Ein}(t)=\int_{0}^{t} (1-\E^{-u}) \D u/u$.

\smallskip
\noindent
{\em Proof.}  The inverse Laplace transform of the first term of (41) which contains the square of logarithmic function can be calculated by using  \cite[Eq. (19.5.2)]{AApelblat13}
\begin{equation}\label{eq28112022-w1}
\mathscr{L}^{-1}[s^{-\lambda - 1} \ln^2 s; t ] = \frac{t^{\lambda}}{\Gamma(\lambda+1)}\big\{[\psi(\lambda+1)-\ln t]^2-\psi'(\lambda+1)\big\}.    \
\end{equation}
Namely, 
\begin{align*}%\label{eq21112022w1}
\mathscr{L}^{-1}\left[\frac{\ln^2 s}{s-1}; t\right] & = \mathscr{L}^{-1}\left[\frac{1}{s}\,\frac{\ln^2 s}{1-1/s}; t\right]
=\mathscr{L}^{-1}\left[(\ln^2 s)\sum_{n=0}^{\infty} s^{-n-1}; t\right]  
 \\ &=\sum_{n=0}^{\infty}\mathscr{L}^{-1}\big[s^{-n-1
 } \ln^2 s; t\big]
=\sum_{n=0}^{\infty}\{[\psi(n+1)-\ln t]^2-\psi^{'}(n+1)\}\frac{t^n}{n!}.
\end{align*}
Hence,
\begin{equation*}%\label{eq21112022w1}
%\begin{split}
\int_0^t \E^{\xi} {\rm Ei}(-\xi) \ln(t-\xi) \D\xi  =\mathscr{L}^{-1}\left[\frac{\ln^2 s}{s-1}; t\right] - \mathscr{L}^{-1}\left[\frac{\ln^2 s}{s}; t \right] + C\, \mathscr{L}^{-1}\left[\frac{\ln s}{s(s-1)}; t\right].
%{\gamma{\mathscr{L}^{-1}\left[\frac{\ln s}{s-1}; t\right]}-\gamma{\mathscr{L}^{-1}\left[\frac{\ln s}{s}; t\right]}}
%\end{split}
\end{equation*}
Using the inverse Laplace transform we have $\mathscr{L}^{-1}\{(\ln s)/[s(s-1)]; t\}=\int_0^t \mathscr{L}^{-1}[(\ln s)/(s-1); u] \D u$, which from Eqs. \eqref{1/09-10} and \eqref{4/09-11} appears to be equal to $\int_0^t M_1(u) \D u = \langle x^2(t) \rangle_1/(2B)$. Now, from Eqs. \eqref{eq6092022-w6} and \eqref{eq28112022-w1} we obtain
\begin{align}\label{eq28112022-w2}
    \begin{split}
\int_0^t \E^{\xi} {\rm Ei}(-\xi) \ln(t-\xi) \D\xi & = \sum_{n=0}^{\infty}\{[\psi(n+1)-\ln t]^2-\psi^{'}(n+1)\}\frac{t^n}{n!}-(\ln t+ C)^2+\frac{\pi^2}{6}\\
& - C \E^{t}{\rm Ei}(-t)+\gamma\ln t+ C^2 \\
& = \sum_{n=0}^{\infty}\{[\psi(n+1)]^2-\psi^{'}(n+1)\}\frac{t^n}{n!}-2(\ln t)\sum_{n=0}^{\infty}\psi(n+1)\frac{t^n}{n!}\\
& + \E^t\ln^2 t-(\ln t+ C)^2 +\frac{\pi^2}{6}- C \E^{t}{\rm Ei}(-t)+C\ln t +C^2.
\end{split}
\end{align}
%\begin{equation}
 %   =\sum_{n=0}^{\infty}\{(\psi(n+1))^2-\psi^{'}(n+1)\}\frac{t^n}{n!}-2(\ln t)\sum_{n=0}^{\infty}\psi(n+1)\frac{t^n}{n!}+e^t\ln^2 t-(\ln t+\gamma)^2.  +\frac{\pi^2}{6}-\gamma \E^{t}{\rm Ei}(-t)-\gamma\ln t-\gamma^2.
% \end{equation}
 Let $H_{n}^{(s)}=\sum_{k = 1}^n k^{-s}$, $s = \mathbb{N}$ is a sequence of harmonic power numbers, where $H_{n}^{(1)} = H_{n}$. Since, $\psi(n+1) = -C + H_{n}$ and $\psi'(n+1) = 
 \pi^2/6 - \sum_{k=1}^n n^{-2} = \pi^2/6 - H_{n}^{(2)}$ , (see \cite {Datoli19, ChS11}),
we obtain,
\begin{align*}%\label{eq28112022-w3}
%\begin{split}
 [\psi(n+1)]^2 -\psi^{'}(n+1) & = C^2 - 2C H_{n}+(H_{n})^2-\frac{\pi^2}{6}+H_{n}^{(2)} \\
 & = -C^2 - 2C\, \psi(n+1) + (H_{n})^2-\frac{\pi^2}{6}+H_{n}^{(2)}.   
%\end{split}
\end{align*}
That allows us to write
\begin{align}\label{eq28112022-w4}
\begin{split}
 \sum_{n=0}^{\infty}\{[\psi(n+1)]^2 -\psi^{'}(n+1)\}\, \frac{t^n}{n!} & - 2(\ln t)\sum_{n=0}^{\infty}\psi(n+1)\frac{t^n}{n!} = -\left(C^2 + \frac{\pi^2}{6}\right) \sum_{n=0}^{\infty} \frac{t^n}{n!} \\ &- 2(C + \ln t)\sum_{n=0}^{\infty} \frac{t^n}{n!}\psi(1 + n) + \sum_{n=0}^{\infty} (H_n)^2\, \frac{t^n}{n!} + \sum_{n=0}^{\infty} H_n^{(2)}\,\frac{t^n}{n!}. 
\end{split}
\end{align}
%For a concise and beautiful description of these numbers, we refer also to Wolfram MathWorld’s Web Site \cite{Wf22}.
Usually it is assumed that $\psi(1) = - C$ which means that $H_0 = 0$. Then, Eq. \eqref{eq28112022-w4} can be presented as
\begin{align*}%\label{eq28112022-w5}
%\begin{split}
  \text{LHS of Eq. \eqref{eq28112022-w4}}
  %\sum_{n=0}^{\infty}\{[\psi(n+1)]^2 -\psi^{'}(n+1)\}& - 2(\ln t)\sum_{n=0}^{\infty}\psi(n+1)\frac{t^n}{n!} 
  & = -\left(C^2 + \frac{\pi^2}{6}\right) \E^{t} + 2 (C^2 + C\ln t) - 2(C + \ln t)\sum_{n=1}^{\infty} \frac{t^n}{n!}\psi(1 + n) \\
  & + \sum_{n=1}^{\infty} (H_n)^2\, \frac{t^n}{n!} + \sum_{n=1}^{\infty} H_n^{(2)}\,\frac{t^n}{n!}. 
%\end{split}
\end{align*}
The abbreviation LHS means left-hand size. Next, we use Eq. \eqref{6/09-5} in which is calculated the generating function for the digamma function and apply results of Ref. \cite{Datoli19}. By employing the umbral calculus the authors show 
%the umbral calculus, Datolli et al. \cite{Datoli19} derived closed form expressions for the generating functions of $H_{n}^{(2)}$ and $(H_{n})^2$:
\begin{align*}%\label{eq28112022-w6}
%\begin{split}
\sum_{n=1}^{\infty} H_{n}^{(2)}\, \frac{t^n}{n!} & = \E^t - 1 + \sum_{n=1}^{\infty} \frac{1}{(n+1)^2} \left(\E^t - \sum_{r=0}^{n}\frac{t^r}{r!}\right) = \frac{\pi^2}{6}(\E^t - 1) - \sum_{n=1}^{\infty}\frac{1}{(n+1)^2}\sum_{r=1}^{n}\frac{t^r}{r!},  \\
\sum_{n=1}^{\infty}(H_{n})^2\, \frac{t^n}{n!} & = \E^t \int_{0}^{t} \big[2\, {\rm Ein}(u) + \E^{-u}\, {\rm Ein}(-u)\big]\, \frac{\D u}{u}    
%\end{split}
\end{align*}
where ${\rm Ein}(s)$ is defined in Proposition \ref{28112022-p1}.
%\begin{equation}
%\sum_{n=1}^{\infty}(H_{n})^2\frac{t^n}{n!}=e^t\int_{0}^{t}\{\frac{2}{s}{\rm Ein}(s)+\frac{e^{-s}}{s}{\rm Ein}(-s)\}\ ds.    
%\end{equation}
%So,
Collecting all obtained in this proof results we get
\begin{align*}%\label{11/11-22}
%\begin{split}
  \text{LHS of Eq. \eqref{eq28112022-w4}} & = -C^2 \E^{t} + 2\E^{t} (C + \ln t)\, \big[{\rm Ei}(-t) - \ln t\big] - \frac{\pi^2}{6} \\
  & - \sum_{n=1}^{\infty} \frac{1}{(n+1)^2} \sum_{r=1}^{n} \frac{t^r}{r!} + \E^t \int_0^t \big[2\, {\rm Ein}(u) - \E^u\, {\rm Ein}(-u)\big] \frac{\D u}{u}.
%\end{split}
\end{align*}
%\begin{align}\label{11/11-22}
%\begin{split}
%\sum_{n=0}^{\infty}\{(\psi(n+1))^2-\psi^{'}(n+1)\}\frac{t^n}{n!} =\gamma e^t (-2\ln t-\gamma+2{\rm E_i}(-t))-\frac{\pi^2}{6} \\ -\sum_{n=1}^{\infty}\frac{1}{(n+1)^2}\sum_{r=1}^{n}\frac{x^r}{r!}+e^t\int_{0}^{t}\{\frac{2}{s}{\rm Ein} (s)+\frac{e^{-s}}{s}{\rm Ein}(-s)\}\ ds.
%\end{split}
%\end{align}
Finally, by substituting the term of the right-hand side of the last equation in Eq. \eqref{eq28112022-w2}, the proof of the proposition is completed. \qed

\end{document}